\documentclass[onecolumn]{aastex631}

\usepackage{amsmath, amssymb}
\usepackage{hyperref} 
\usepackage{epsfig}

\usepackage{placeins}
\usepackage{wrapfig}

\def\fluxthres{\hat f_{\bar \e}}

\def\zmax{z_{\rm max}}

\def\e{\epsilon}

\def\Estarg{{\cal E}_{*\gamma}}

\def\Swift{\emph{Swift}}

\def\H0{H_{\rm 0}}

\newcommand{\begeq}{\begin{equation}}
\newcommand{\fineq}{\end{equation}}
\newcommand{\begfig}{\begin{figure}}
\newcommand{\finfig}{\end{figure}}
\newcommand{\begeqarray}{\begin{eqnarray}}
\newcommand{\fineqarray}{\end{eqnarray}}

\makeatletter
\def\@journalinfo{Accepted for publication to MNRAS}
\def\@printdate{}
\makeatother

\shorttitle{GRBs Distributions and the Hubble Constant $\H0$} 
\shortauthors{Le and Chromicz}

\begin{document}

\title{The LGRBs Redshift and Jet Opening Angle Distributions and the Hubble Constant $\H0$}

\author[0000-0000-0000-0000]{Truong Le}
\affiliation{Department of Mathematics, Computer Science, and Physics, Roanoke College, Salem, VA 24153, USA; tle@roanoke.edu }

\author[0000-0000-0000-0000]{Joseph Chromicz}
\affiliation{Department of Physics, American University, Washington, DC 20016, USA}



\begin{abstract}
Using the \Swift-Perley sample, Le, Ratke, \& Mehta performed a study of the rate density of LGRBs with an assumed broken power-law GRB spectrum. They obtained a GRB-burst-rate functional form that fits the pre\Swift~and \Swift~redshift cumulative distributions at all redshifts with an assumed Hubble constant of $\H0 = 72 \, {\rm km \, s^{-1} \,  Mpc^{-1}}$. Further analysis indicated that the cumulative redshift distribution is weakly sensitive to variations in the Hubble constant $\H0$, whereas the corresponding differential burst-rate distribution exhibits greater sensitivity. In this paper, we extend this analysis by comparing model predictions for different values of $\H0$ with the observed cumulative and differential redshift and jet-opening angle distributions, while adopting the same GRB framework and observational samples. We find that both the cumulative and differential distributions depend only weakly on variations in $\H0$, and the model predictions using $\H0 \approx 68 - 80 \,  {\rm km \, s^{-1} \,  Mpc^{-1}}$ provide agreement with the observed LGRB distributions.
\end{abstract}


\section{Introduction} \label{sec:intro}

The Hubble constant, $\H0$, has been a cornerstone of cosmology since its introduction by Edwin Hubble in 1929. The characteristic of this value reveals the current expansion rate and the observable size of the Universe, and its inverse sets the expansion age of the Universe. We have used this value to understand the nature of supermassive black holes, gamma-ray bursts (GRBs), particle physics, and most importantly, the relative amount of matter, dark matter, and dark energy in the universe. Any changes to the value of $\H0$ would alter our understanding of the universe’s age and the balance of these components.

Using the standard-candle method, which is based on the inverse square law of flux giving an estimate of the distance to a light source, and the temperature and polarization fluctuations in the cosmic microwave background (CMB), the current Hubble constant is estimated to be $73.24 \pm 1.74 \, {\rm km \, s^{-1} \, Mpc^{-1}}$~\citep[][]{rie16} and $67.8 \pm 0.9 \, {\rm km \, s^{-1} \,  Mpc^{-1}}$~\citep[][]{ade16}, respectively.  Besides the standard-candle and CMB methods, GRBs have also been proposed as an independent probe of the Hubble constant because of their extreme luminosities and broad redshift coverage. These studies estimate $\H0$ by calibrating empirical GRB luminosity correlations (e.g., Amati-, Yonetoku-, and related relations), constructing GRB Hubble diagrams, and fitting the resulting luminosity distance - redshift relation to cosmological models \citep[e.g.,][]{ad13,wan16,des21,lm21,dvd24,lyy24,wl24,bdc25}.

Briefly, GRBs are brief flashes of gamma-rays occurring at an average rate of a few per day throughout the universe. The ultimate energy source of a GRB is believed to be associated with a catastrophic event associated with black-hole formation \citep{mr97}. This event takes place through the collapse  of the core of a massive star in the case of long-duration GRBs (LGRBs), and due to merger or accretion-induced collapse events for the short-hard class of GRBs. Since LGRBs are associated with the deaths of massive stars, it is commonly assumed that the GRB rate density follows the observed star formation rate (SFR) density history \citep[e.g.,][]{pac98}. Because of their high luminosity, up to $10^{54}$ erg s$^{-1}$ \citep[see][]{pes16}, GRBs can be detected out to the early universe \citep[e.g.,][]{bl06}, and the farthest GRB to date is GRB 090429B with a photometric redshift $z = 9.4$ \citep[][]{cuc11}. Hence, this holds a promise that GRBs can be used to probe the early universe.

With the launch of the Swift satellite, rapid follow-up observations of gamma-ray bursts (GRBs) triggered by the Burst Alert Telescope became possible, leading to the discovery of a fainter and more distant GRB population than had been detected by earlier missions such as BATSE, BeppoSAX, INTEGRAL, and HETE-2. The mean redshift of the 41 long-duration GRBs (LGRBs) in the pre\Swift~sample is $\langle z \rangle \approx 1.5$ \citep[][hereafter pre\Swift-Friedman]{fb05}, whereas the 16 GRBs in the initial~\Swift~sample have a mean redshift of $\langle z \rangle \approx 2.72$ \citep[][hereafter~\Swift-Jakobsson]{jak06}. Using these early samples, \citet[][]{bag06} showed that the~\Swift~and pre\Swift~redshift distributions are statistically different, suggesting that they probe different subsets of the overall GRB population because of differences in detector sensitivity and follow-up capabilities. To understand these differences, \citet[][]{ld07} and \citet[][]{lm17} developed a phenomenological GRB framework that incorporates detector triggering effects, GRB spectral properties, and the evolution of the LGRB comoving rate density to reproduce the observed redshift and jet opening-angle distributions of both the early and later, more complete, pre\Swift~and~\Swift~samples. 

Most recently, using the pre\Swift-Friedman sample together with the more complete and reliable~\Swift-Perley redshift sample and the~\Swift-Ryan jet opening-angle sample, \citet[][]{lrm20} showed that the phenomenological GRB framework also provides acceptable fits to the observed cumulative redshift and jet opening-angle distributions for an assumed Hubble constant of $H_0 = 72$ km s$^{-1}$ Mpc$^{-1}$. Most importantly, they found that the cumulative redshift distributions are only weakly sensitive to variations in the Hubble constant. However, when the same information is expressed as directional differential event-rate distributions, $d{\dot{N}}/d\Omega dz$ and $d{\dot{N}}/d\Omega d\mu$, the sensitivity to $\H0$ becomes stronger because differential distributions preserve local variations that are smoothed out in cumulative representations. Since many GRBs studies fix $\H0$ near $68-73 \, {\rm km \, s^{-1} \,  Mpc^{-1}}$, the demonstrated sensitivity of the $d{\dot{N}}/d\Omega dz$ or $d{\dot{N}}/d\Omega d\mu$ could imply a real risk of parameter bias in the inferred LGRB rate density evolution. Hence, in this paper, we revisit the analysis of \citet[][]{lrm20} and treat the local Hubble constant $\H0$ as a free parameter. To ensure a direction comparison with their paper, we adopt the same GRB framework (see Tables 1, 2, and 3) and observational samples (see Tables 4, 5, and 6) that were previously shown to provide the best fits to the observed redshift and jet opening-angle distributions. Within this framework, we adopt the previously established LGRB formation-rate evolution models and examine how variations in $\H0$ affect the predicted differential event-rate distributions.

For clarification, in our previous papers and in this work, the term fit, good fit, acceptable fit, and best fit refer to varying the model parameters to obtain agreement between the predicted and observed distributions. The Kolmogorov–Smirnov (KS) test is then applied specifically to the cumulative redshift and jet opening-angle distributions to assess their statistical consistency with the observed samples. Thus, our approach involves model fitting through parameter exploration rather than a systematic optimization procedure based on minimizing a statistic such as \(\chi^2\) or maximizing a likelihood function. This procedure is discussed further at the end of Section~4. 

In Sections~2 and 3, we briefly summarize key results and review the pre\Swift~and~\Swift~LGRB samples that were in \citet[][]{lrm20}. We also summarize the observed directional differential event-rate distributions derived from these samples, which form the basis of the analysis presented in this paper. In Section~4, we briefly summarize the \citet[][]{ld07} cosmological GRB framework, the redshift and the jet opening angle distributions equations, and our fitting procedure. In Section~5, we discuss the results, and conclude with a discussion of the inferred local Hubble constant in Section~6. Throughout, we assume a flat $\Lambda$CDM cosmology with $\Omega_{m} = 0.27$ and $\Omega_{\Lambda} = 0.73$ \citep[e.g.][]{spe03} while varying the local Hubble constant $\H0$. 

\section{AN OVERVIEW OF THE LRM20 RESULTS}
\citet[][]{ld07}, \citet[][]{lm17}, and \citet[][]{lrm20} developed a phenomenological GRB framework that combines detector triggering effects, GRB spectral properties, jet opening angle distribution function, and the evolution of the LGRB comoving rate density to explain the observed redshift and jet opening angle distributions across the pre\Swift~and~\Swift~samples. The framework adopts a uniform jet geometry together with a flat or broken power-law $\nu F_\nu$ spectrum and assumes that the LGRB rate evolution broadly follows the observed cosmic star-formation history. The two leading GRB jet frameworks are (1) the uniform jet model, in which the energy per solid angle is approximately constant within a well-defined jet opening angle $\theta_i$, but drops sharply outside of $\theta_i$, with differing GRBs having different jet opening angles \citep[see][]{mr97,mrw98,fra01,ber03,ggl04,sal14,pgs15,bn19,lha20}; and (2) the universal structured jet model, in which all GRB jets are intrinsically similar and the directional energy release decreases approximately as the inverse square of the opening angle from the jet axis \citep[][]{lpp01,rlr02,zm02,zha04,sal15,sg22,gts25}. 

While alternative jet structures have been proposed, including quasi-universal, power-law, and hybrid angular jet models \citep[][]{rlr02,zm02,zha04,sal15,sg22,gts25}, we adopt the uniform jet model in this work for several reasons. First, the uniform jet model provides a simple phenomenological description with a well-defined jet opening angle that can be directly compared with the observed jet opening angle distributions used in this analysis. Second, previous studies \citep[][]{ld07,lm17,lrm20} showed that this framework successfully reproduces the observed cumulative redshift and jet opening angle distributions of the pre\Swift~and~\Swift~samples while accounting for detector triggering effects and selection biases. Finally, the primary goal of this paper is not to determine the true physical structure of GRB jets, but rather to investigate how variations in the Hubble constant affect the differential event-rate distributions within the same framework adopted in previous studies. Therefore, to maintain consistency with the \citet[][]{lrm20} analysis, we continue to adopt the uniform jet model in the present work. Exploring alternative jet structures is beyond the scope of this paper.

Evidence for jetted GRBs comes from achromatic breaks observed in radio and optical afterglow light curves \citep[][]{wkf98,sta99}. Subsequent observations with Swift/XRT and later Chandra follow-up studies showed that jet breaks may occur at much later times than initially expected, enabling estimates of GRB jet opening angles \citep[][]{lia08,rac09,zha15}. Using these methods, \citet[][]{zha15} and \citet[][]{rya15} derived jet opening-angle measurements for Swift GRBs that forms the basis of the jet opening angle analysis in both \citet[][]{lrm20} and the present work. Although the uniform jet interpretation has been challenged by the complex afterglow behavior revealed by Swift observations, jet breaks continue to be studied and interpreted within both uniform and structured jet frameworks \citep[e.g.,][]{rac09,wan18,lam21}.

\begfig[t] \hskip-0.3in \vskip-0.15in \epsscale{1.0} \plotone{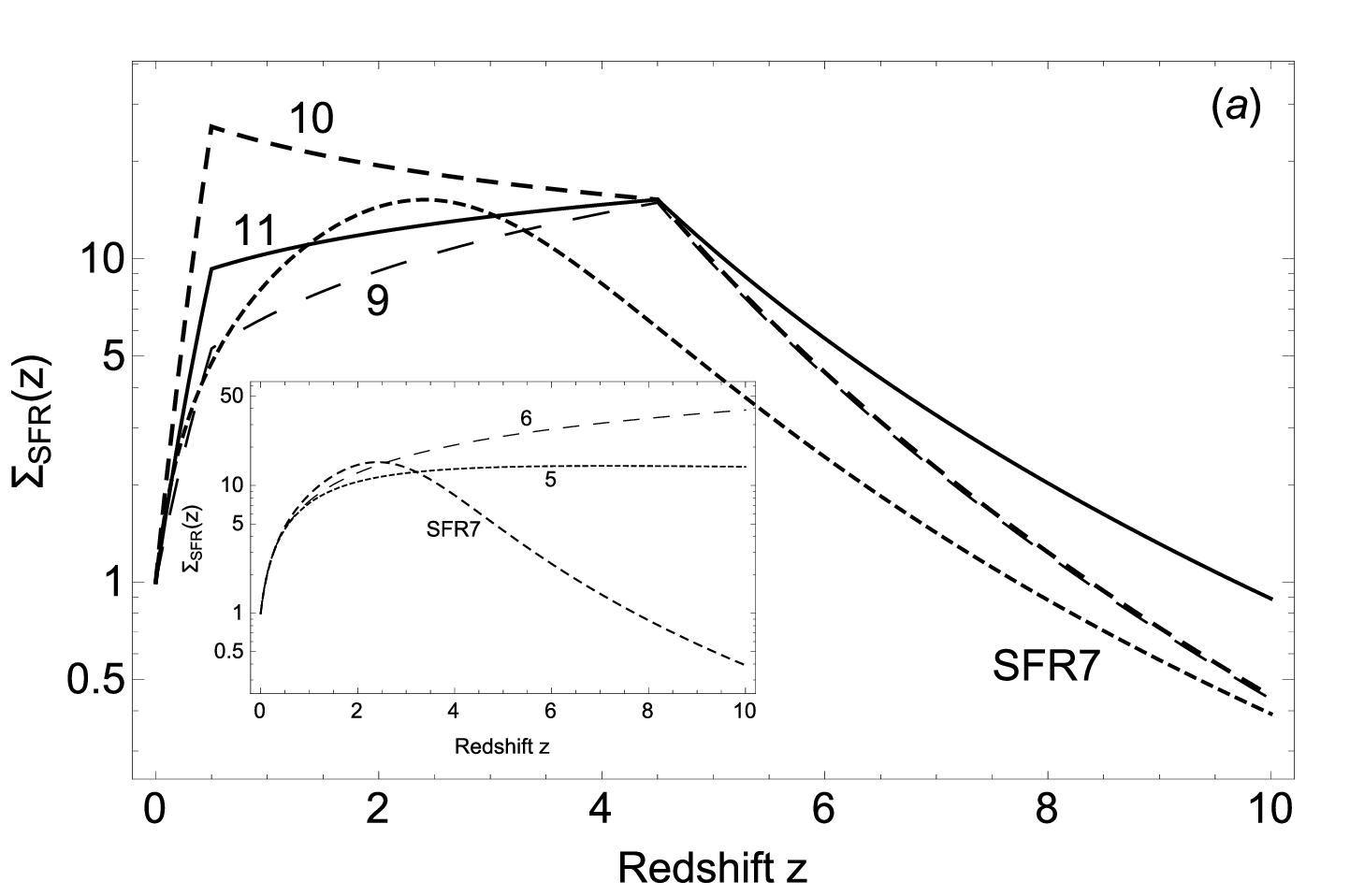}
\caption{\footnotesize (a) These curves represent the GRB density-rate models from \citet[][]{ld07}, \citet[][]{lm17}, and \citet[][]{lrm20}. Curves SFR5 and SFR6 (inset) provide acceptable fits to the pre\Swift-Friedman and~\Swift-Jakobsson cumulative redshift and jet opening angle distributions assuming a flat GRB spectrum. Curve SFR9 provides acceptable fits to the pre\Swift-Friedman and~\Swift-Ryan cumulative distributions assuming a broken power-law GRB spectrum. SFR7 represents the observed cosmic star-formation rate history from \citet{hb06}, extended by~\citet{li08}. Curves SFR10 and SFR11 provide acceptable fits to pre\Swift-Friedman,~\Swift-Perley (redshift) and~\Swift-Ryan (jet opening angle) samples, with SFR11 providing the preferred fit. (b) Cumulative redshift distributions of 133 GRBs in the Swift-Ryan sample (thick dark curve) and the pre\Swift-Friedman sample (inset; thick dark curve) from \citet[][]{lm17}.} 
\label{fig1} 
\finfig
\begfig[t] \hskip-0.25in \epsscale{1.2} \plotone{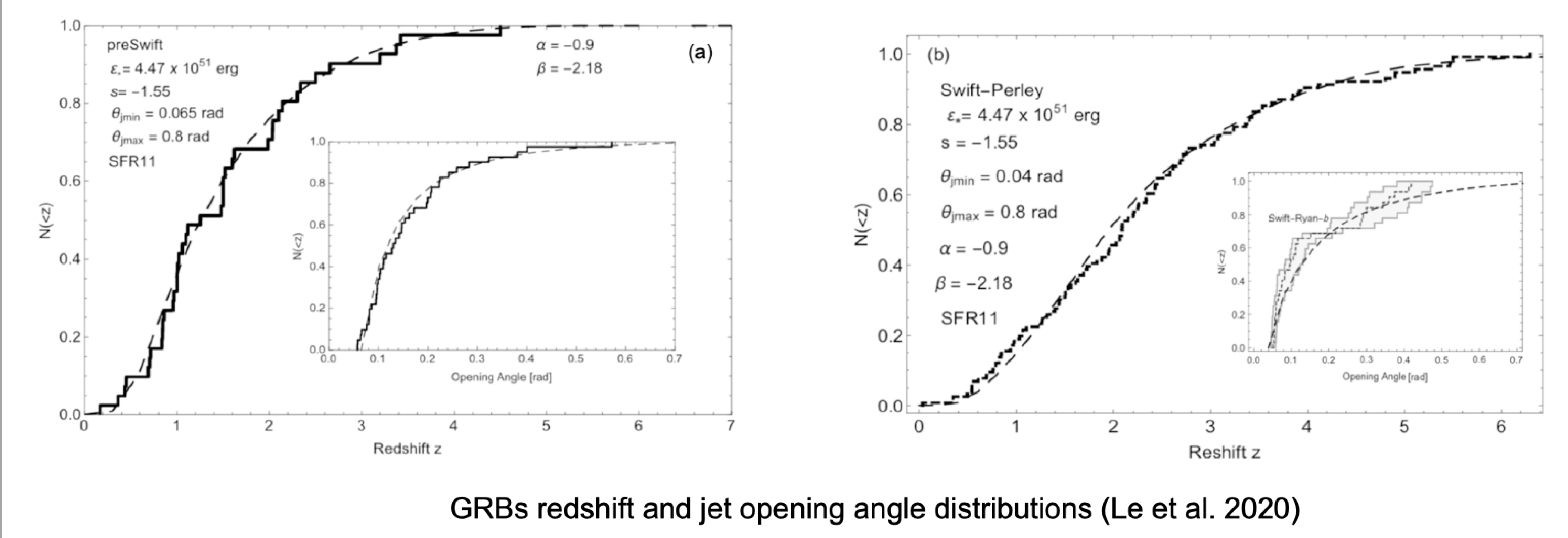}
\caption{\footnotesize ({\it a}) The pre\Swift~cumulative redshift and jet opening angle distributions of Figure~7(b) from \citet{lrm20}. (b) The \Swift-Perley {\bf cumulative} redshift distribution of Figure~8(b) and the \Swift-Ryan-b jet opening angle distribution of Figure~9(a) from \citet{lrm20}. In both panels, the long-dashed curves represent the model predictions, while the thick dark curves and shaded gray regions represent the observed distributions, respectively.}
\label{fig2} 
\finfig

\citet[][hereafter LD07]{ld07} assumed a flat $\nu F_\nu$ GRB spectrum as their initial study to understand the difference between the redshift and jet opening angle distributions of the initial samples (pre\Swift-Friedman and~\Swift-Jakobsson) due to different detector characteristics, but a more realistic model is a broken power-law $\nu F_\nu$ GRB spectrum, which was later studied by \citet[][]{lm17} as discussed below. LD07 assumed that the LGRB density rate is proportional to the measured star formation rate (SFR), where SFR7 in the inset of Figure~\ref{fig1} is the observed \citet{hb06} SFR history and later extended by \citet[][]{li08}; Table~\ref{tbl-1} summarizes the GRB formation-rate model parameters associated with different star-formation rate histories and Table~\ref{tbl-2} represents the pre\Swift~and~\Swift~detectors sensitivity, which we further discuss in Sections~3 and 4. However, Le \& Dermer showed that a good fit was only possible by providing a positive evolution of the SFR history of GRBs to high redshifts that is similar to SFR5 or SFR6, see the inset of Figure~\ref{fig1}. Their calculated LGRB density rate models, SFR5 and SFR6, at high redshift were consistent with other researchers \citep[e.g.][]{lfr02,dai06,mes06,gp07,kis08,wp10}. However, it was unclear whether the excess at high redshift was due to luminosity evolution \citep[e.g.,][]{sc07,sal09,den16} or the cosmic evolution of the GRB rate \citep[e.g.,][]{but10,qin10,wp10}. Furthermore, \citet[][]{sal12}, for example, suggested that a broken power law luminosity evolution with redshift is required to fit the observed redshift distribution. 

\citet[][hereafter LM17]{lm17} revisited the work done by LD07 and performed a timely study of the rate-density of GRBs with an assumed broken power-law GRB spectrum with the low- and high-energy indices ($\alpha = -1, \beta = -2.5$) of the Band function \citep[][]{ban93}. Utilizing more than 100 LGRBs in the Swift sample that include both the observed estimated redshifts and jet opening angles \citep[][hereafter~\Swift-Ryan-2012 sample, also see Table~5]{zha15,rya15}, LM17 obtained a GRB burst rate functional forms, SFR9 and SFR10 in Figure 1, that produced the model distributions that were comparable to the observed pre\Swift-Friedman and~\Swift-Ryan-2012 redshift distributions (see their Figures 5, 7 and 8), with SFR10 providing the preferred fit. The one-sample Kolmogorov–Smirnov (KS-1) test indicates that the observed estimated~\Swift-Ryan and pre-Swift redshift and jet opening angle distributions and their associated fit probability statistics (p statistics) are greater than 0.05, so the null hypothesis is rejected, indicating that the samples and their associated fits belong to the same distribution. They also showed that the SFR5 and SFR6 models that were utilized by LD07 could not fit the more current updated LGRB~\Swift-Ryan-2012 sample. Interestingly, SFR9 and SFR10 are similar to the \citet[][]{hb06} star formation history (SFR7) and as extended by \citet[][]{li08}. Most importantly, however, the result indicated an excess of LGRBs at low redshift below $z \sim 2$ in the Swift sample (see their Figure 7(a)), consistent with other researchers \citep[e.g.,][]{pkk15,yu15,laj19}. Nevertheless, the reason for this excess was either unclear, incomplete sample size, or that GRB formation rate did not trace SFR at low redshift less than $z \le 1$ \citep[e.g.,][]{pkk15,yu15,laj19}. 

\citet[][hereafter LRM20]{lrm20} revisited the work done by LM17 addressing the excess of LRGB at low redshift below $z \sim 2$ in the~\Swift-Ryan-2012 and~\Swift-Perley samples. They  demonstrated that the~\Swift-Ryan sample exhibits an excess of LGRB at low redshift ($z < 2$), whereas the~\Swift-Perley sample (see Table~6) does not (see their Figures~5(b), 6(b), 8(a)-8(b)), with SFR11 providing the preferred GRB burst rate model. Their result suggested that the low-redshift excess is primarily due to selection effects rather than an intrinsic feature of the GRB population, which we will further discuss in Section~3. The~\Swift-Perley sample provides reliable redshift data but does not contain jet opening angle data, while the~\Swift-Ryan sample includes jet opening angle measurement. LRM20 combined the pre\Swift-Friedman sample and the~\Swift-Perley redshift and the~\Swift-Ryan jet opening angle samples to constrain the GRB model parameters presented here in Tables 1-2, and Case-1 and Case-2 in Table~3.

Using the pre\Swift-Friedman (redshift and jet opening angle) sample and the~\Swift-Perley (redshift) and the~\Swift-Ryan (jet opening angle) samples, LRM20 demonstrated that their GRB model provide a good fit to both the cumulative redshift and jet opening angle distributions, (for example, see their Figures 7(b), 8(b), and the inset of 9(a), for convenient we also presented them here in Figures~2(a) and 2(b)), and that the inferred LGRB density rate (SFR11, see their Figure 1(a)) generally follows the observed cosmic star-formation rate for an assumed Hubble constant of $\H0 = 72 \, {\rm km \, s^{-1} \,  Mpc^{-1}}$. They also showed that the Kolmogorov-Smirnov two-sample test of the pre\Swift-Friedman and the~\Swift-Perley redshift distributions yielded values of approximately $0.11$ for the D-statistic and $0.39$ for the p-value, indicating that the observed distributions are statistically consistent with being drawn from the same parent distribution. Additionally, the one-sample Kolmogorov-Smirnov test indicated that the observed estimated pre\Swift-Friedman and~\Swift-Perley redshift and~\Swift-Ryan jet opening angle distributions and their associated fit probability statistics ($p$ statistics) are greater than $0.05$, so the null hypothesis is rejected. Most importantly, they found that the redshift distribution model is largely insensitive to variations in the Hubble constant ($\H0 = 68,70,72 \, {\rm km \, s^{-1} \,  Mpc^{-1}}$; for example, see their Figure 3(a), for convenient we also presented it here in Figure~3(a)) when expressed as a cumulative distribution function. However, the model becomes sensitive to $\H0$ when the distribution is represented as a directional differential event rate per redshift, $d{\dot{N}}/d\Omega dz$ (for example, see their Figure 2(b), for convenient we also presented it here in Figure~3(b)). The reason for the insensitivity is because the cumulative redshift distribution smooths the data by integrating all bursts up to a given redshift, which can hide subtle differences between models with different $\H0$ values. In contrast, the differential burst-rate per redshift distribution preserves the local structure and variation of the data within each redshift bin. As a result, changes in $\H0$ more directly affect the shape, peak location, and normalization of the differential distribution, making it more sensitive to variations in the Hubble constant. Motivated by this greater sensitivity, the present work explores how variations in $\H0 = 68,73,80 \, {\rm km \, s^{-1} \,  Mpc^{-1}}$ affect the predicted differential burst-rate distributions.

\begfig[t] \hskip-0.25in \epsscale{1.2} \plotone{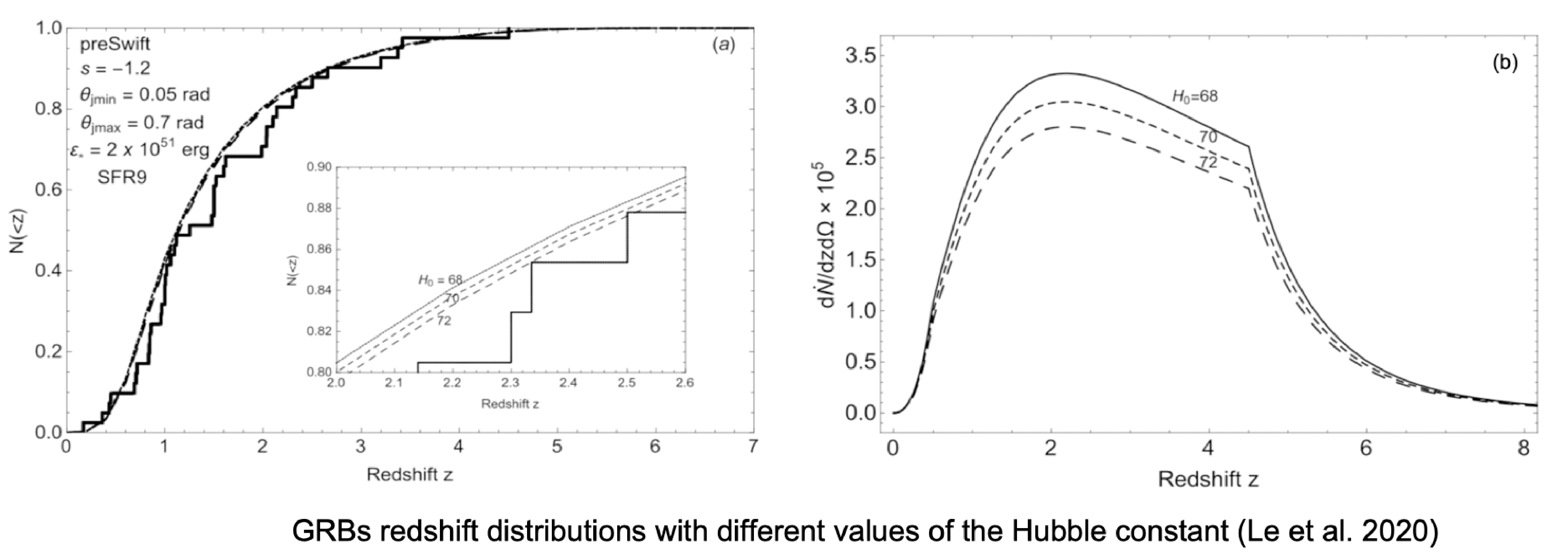}
\caption{\footnotesize ({\it a}) The calculated and the observed pre\Swift~accumulative redshift distribution with different values of the Hubble constant in Figure~3a from \citet{lrm20} paper. (b) The calculated differential burst rate per redshift distribution with different values of the Hubble constant in Figure~2a from \citet{lrm20} paper.}
\label{fig3} 
\finfig
\begfig[t] \hskip-0.25in \epsscale{1.1} \plottwo{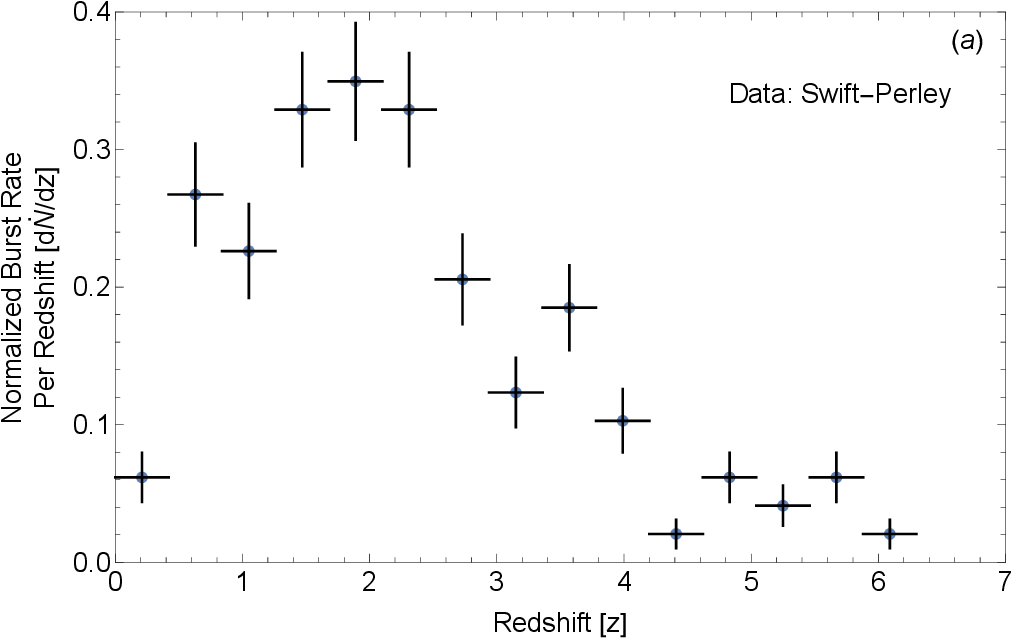}{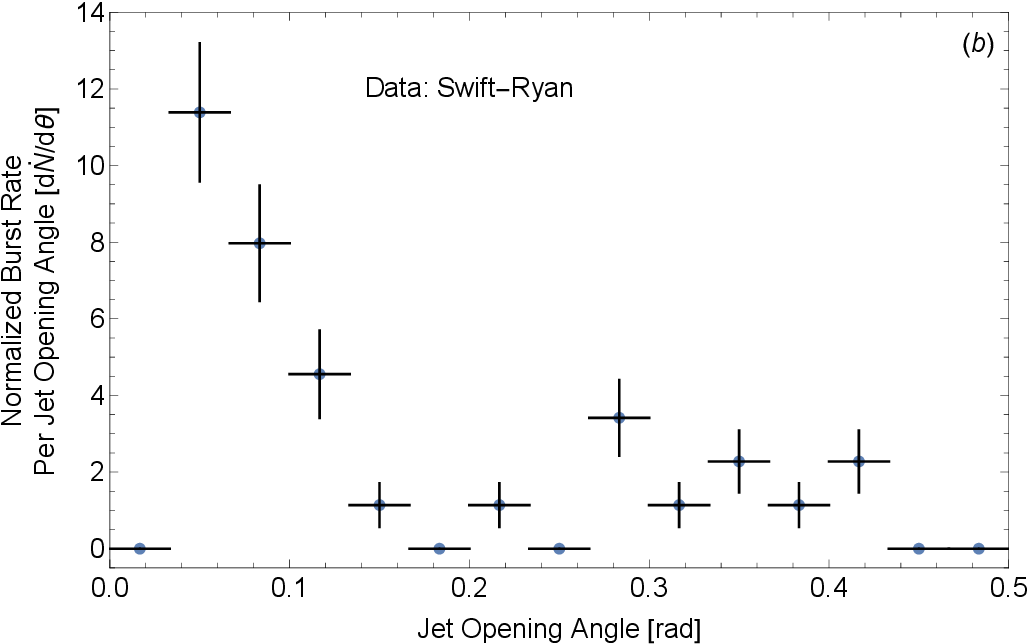}
\caption{\footnotesize ({\it a}) The normalized observed differential burst rate per redshift distribution and ({\it b}) the normalized observed differential burst rate per angle distribution from the~\Swift-Perley and~\Swift-Ryan data as shown in Tables~\ref{tbl-4} and \ref{tbl-5}. The crosses of each data point along the x-axis and y-axis represent the bin size and the statistical error per bin from their samples. The five zero-value burst data points in Figure (1b) indicate that no bursts were observed within those jet opening angle bins. Therefore, these bins can be treated as having no detected data points.}
\label{fig4} 
\finfig
\begfig[t] \hskip-0.25in \epsscale{1.1} \plottwo{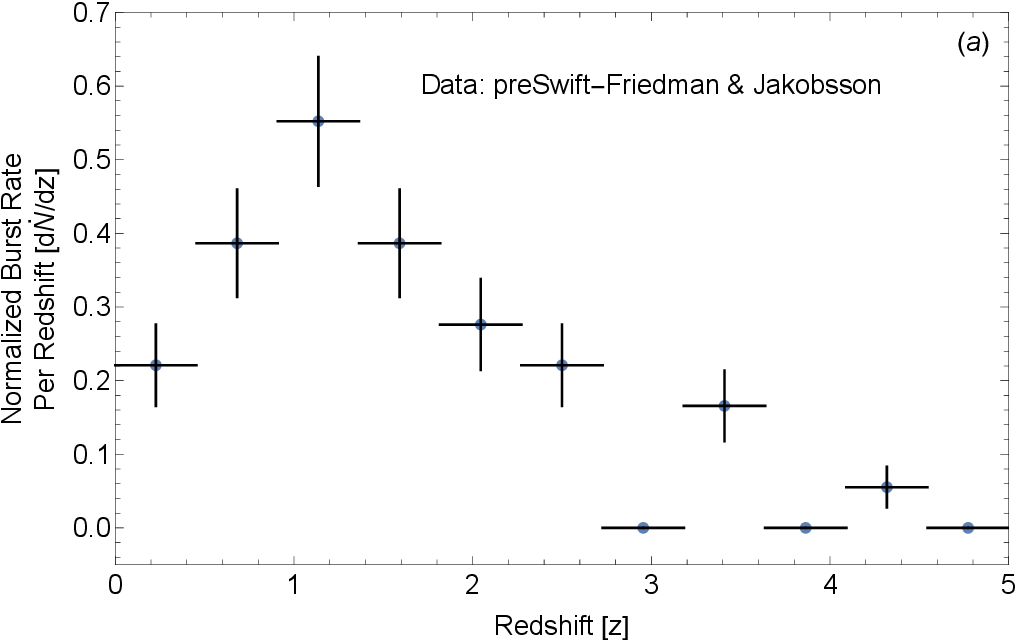}{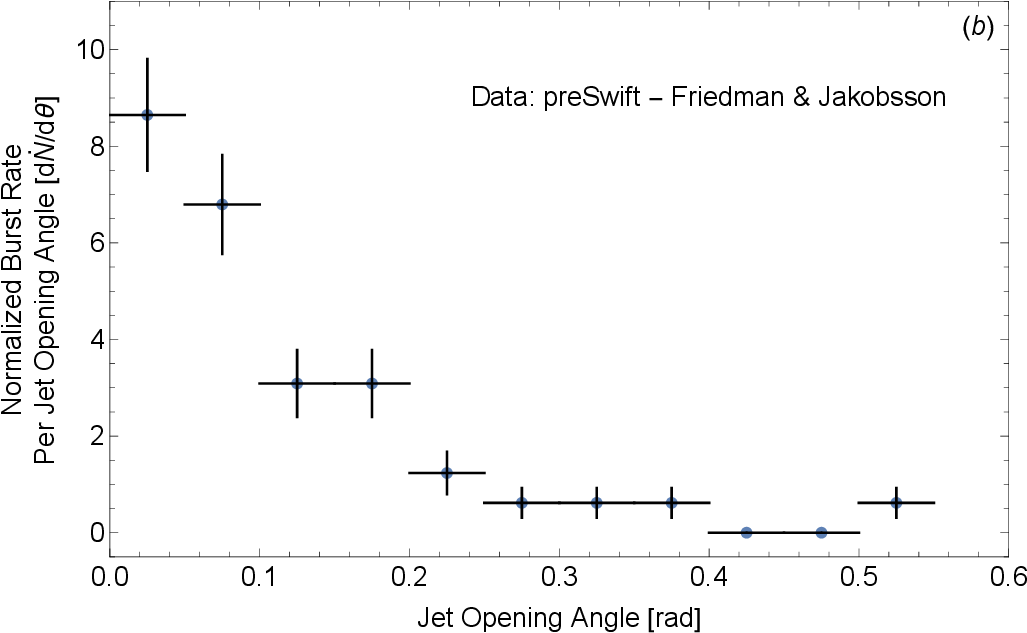}
\caption{\footnotesize ({\it a}) The normalized observed differential burst rate per redshift distribution and ({\it b}) the normalized observed differential burst rate per angle distribution from the pre\Swift-Friedman \& Jakobsson data as shown in Tables~\ref{tbl-3}. The crosses of each data point along the x-axis and y-axis represent the bin size and the statistical error per bin from their sample. The three zero-value burst data points in Figure (2a) and two zero-value burst data points in Figure (2b) indicate that no bursts were observed within those redshift or jet opening angle bins. Therefore, these bins can be treated as having no detected data points.}
\label{fig5} 
\finfig
\begfig[t] \hskip-0.25in \epsscale{1.1} \plottwo{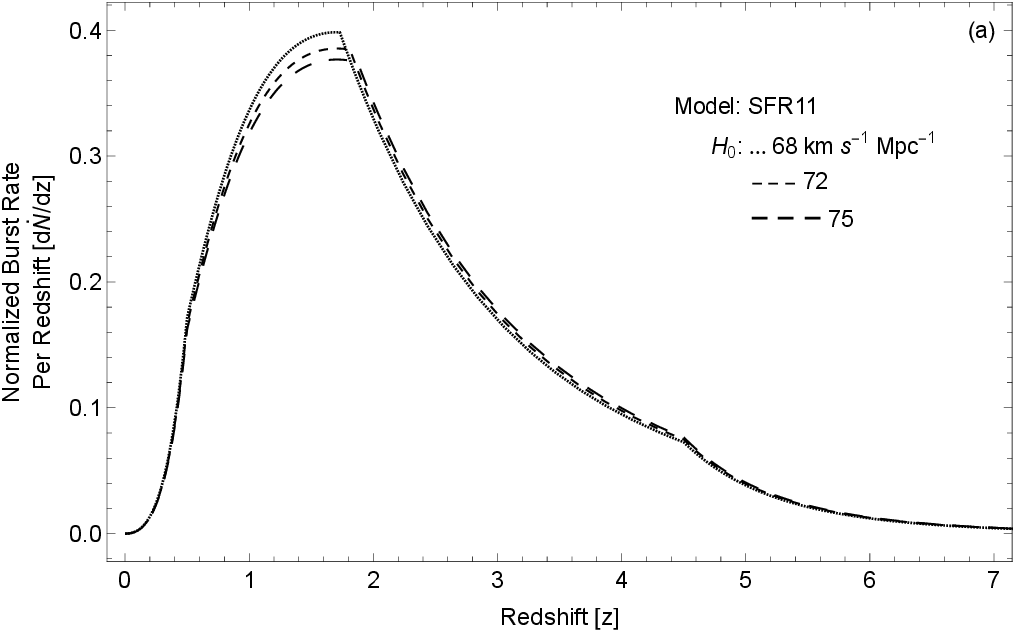}{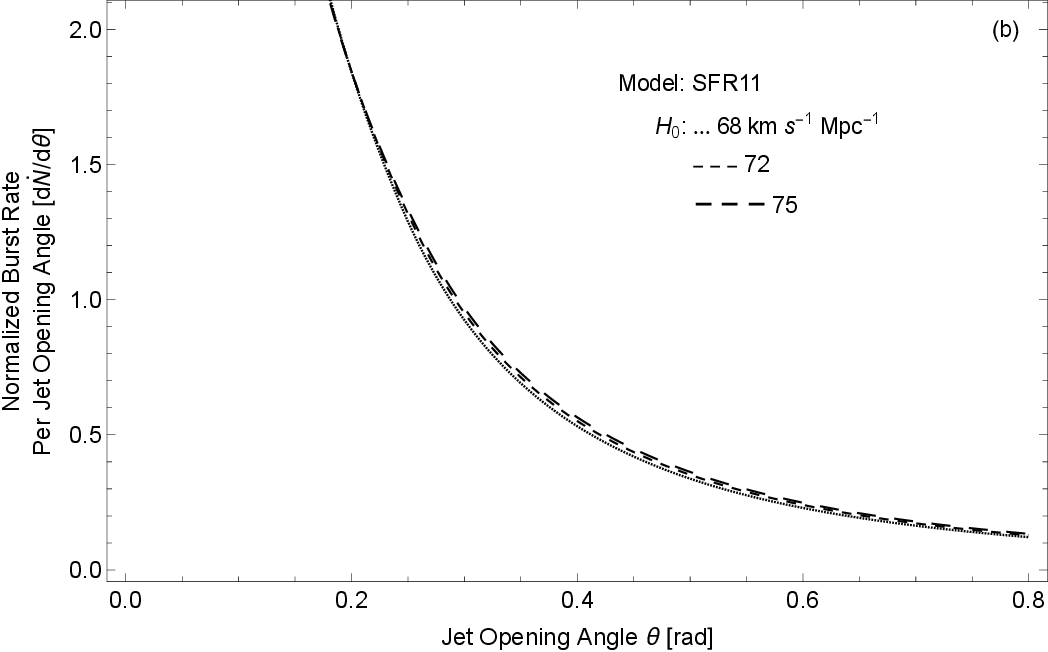}
\caption{\footnotesize ({\it a}) The normalized event rate per unit redshift [one event per $(10^{28} \ \rm cm)^3$ per day per $z$] and ({\it b}) the normalized event rate per unit jet opening angle [one event per $(10^{28} \ \rm cm)^3$ per day per $\theta_{\rm j}$] using $\theta_{\rm j,min} = 0.04 \ \rm rad$ and $\theta_{\rm j,max} = 0.8 \ \rm rad $,  $s = -1.55$, $\Estarg = 4.47 \times 10^{51} \ \rm erg$, $\fluxthres = 10^{-8} \ \rm erg \ cm^{-2} \ s^{-1}$,  $\alpha = -0.9$, $\beta = -2.18$, and SFR11. The solid, short, and long dash curves represent the Hubble constant values of $H_0 = 68, 72$, and $75$ km s$^{-1}$ Mpc$^{-1}$, respectively.}
\label{fig6} 
\finfig

\section{A BRIEF SUMMARY OF THE {\it PRE-SWIFT} AND {\it SWIFT} SAMPLES \\ AND THEIR DIFFERENTIAL BURST-RATE DISTRIBUTIONS}
Table~\ref{tbl-1} summarizes the GRB formation-rate model parameters associated with LD07, LM17, and LRM20 papers depending on which~\Swift~samples were utilized in their studies. For example, LD07 used the GRB formation-rate models SFR5 and SFR6 to explain the initial pre\Swift-Friedman and~\Swift-Jakobsson redshift and jet opening angle samples, while LRM20 used the GRB formation-rate models SFR7, SFR9, and SFR11 to explain the pre\Swift-Friedman and the more later and complete complete~\Swift~samples (\Swift-Perley and~\Swift-Ryan). Table~\ref{tbl-2} lists the triggering sensitivities and detection thresholds of the instruments used in the analysis for both the pre\Swift~and~\Swift~samples. Following LD07, LM17, LRM20 and references therein, we adopt the energy flux $10^{-7}$ erg cm$^{-2}$ s$^{-1}$ and $10^{-8}$ erg cm$^{-2}$ s$^{-1}$ as the effective pre\Swift~and~\Swift~detection flux thresholds, respectively. 

Table~\ref{tbl-3} presents the adopted GRB physical model parameters in LRM20, including quantities related to the jet opening angle distribution $(\theta_{j,min}, \theta_{j,max}, s)$, gamma-ray energy release $(\Estarg)$, and spectral assumptions $(\alpha, \beta)$. Case-1 and Case-2 represent the parameter sets that provide agreement with the cumulative redshift and jet opening angle for the  pre\Swift-Friedman, the~\Swift-Perley redshift, and the~\Swift-Ryan redshift jet opening angle samples. The primary difference between these two cases is the minimum opening jet angle ($\theta_{j,min}$). To obtain the closet agreement with the pre\Swift-Friedman, and the~\Swift-Perley redshift and~\Swift-Ryan redshift jet opening angle samples, the model requires $\theta_{j,min} \sim 0.065$ rad for the pre\Swift~and $0.04$ rad for the~\Swift. Using these parameters, the model predictions were compared with the observed cumulative redshift and jet opening angle distributions of the LGRBs pre\Swift~sample(Table~\ref{tbl-4}) and the~\Swift-Perley redshift sample (Table~\ref{tbl-5}) and the~\Swift-Ryan jet opening angle sample (Tables~\ref{tbl-6}). The corresponding results are shown in Figures 7(b), 8(b) and the inset of 9(a) of LRM20 and those figures are also present here in Figures~\ref{fig2}(a)-\ref{fig2}(b) as we have discussed above. 

Table~\ref{tbl-4} contains the pre\Swift-Friedman sample compiled from \citet[][]{fb05}, which relied heavily on ground-based optical follow-up for redshift determination, and it is subject to several observational selection effects, including incomplete redshift measurements. 
Table~\ref{tbl-5} contains the~\Swift-Perley redshift sample from \citet[][]{per16}, which benefited from the rapid localization capability of the Swift satellite and more systematic afterglow follow-up observations. \citet[][]{per16} constructed the Swift Gamma-Ray Burst Host Galaxy Legacy Survey (SHOALS), a large and relatively unbiased Swift LGRB sample designed to reduce observational biases associated with unfavorable ground-based follow-up conditions. This effort is similar to the Optically Unbiased GRB Host (TOUGH) Survey \citep[][]{hjo12,jak12,rya15}, which obtained 90\% completeness for 69 LGRBs; we refer to this as the~\Swift-Ryan sample (see Table~\ref{tbl-6}). Compared to the earlier~\Swift-Ryan sample, the Swift-Perley sample addresses additional biases affecting redshift determination and includes bursts with bright afterglow observations, making it a more reliable sample for studying the intrinsic LGRB redshift distribution. By applying observability cuts and improving redshift recovery, the SHOALS sample achieved about 92\% redshift completeness for 110 LGRBs. 

It is important to emphasize again that LRM20 used the pre\Swift-Friedman,~\Swift-Perley (redshift), and~\Swift-Ryan (jet opening angle) samples (see Tables~\ref{tbl-4}-\ref{tbl-6}) to constrain their GRB physical model parameters listed in Tables~\ref{tbl-1}-\ref{tbl-2} and Case-1 and Case-2 in Table~\ref{tbl-3}. In the present work, we adopt the same observational samples and GRB model parameters to examine how variations in the Hubble constant affect the predicted directional differential event-rate distributions in redshift and jet opening angle. In Figures~\ref{fig4}(a)-\ref{fig4}(b) and \ref{fig5}(a)-\ref{fig5}(b), we plot the observed differential burst rate per redshift and the observed differential burst rate per angle distributions for both the pre\Swift~(Friedman \& Jakobsson)~and~\Swift~(Ryan \& Perley)~samples based on the redshift and jet opening angle data listed in Tables~\ref{tbl-4}, \ref{tbl-5}, and~\ref{tbl-6}. The distributions are normalized such that the sum of the burst counts over all bins equals unity, thereby representing the relative probability distribution of bursts as a function of redshift or jet opening angle rather than the absolute number of detected events. This normalization facilitates direct comparisons between samples of different sizes and between the observations and model predictions. The zero-value data points shown in Figures \ref{fig4}(b) and \ref{fig5}(a)-\ref{fig5}(b) are not physical measurements; rather, they indicate redshift or jet opening angle bins in which no bursts were observed. Although these points could be removed from the plots, we retain them to explicitly show the absence of detected bursts within those bins.

In this work, we use only the jet opening angle data from the \Swift-Ryan sample; however, for completeness, we also include the associated redshift data in Table~\ref{tbl-6}. The redshift and jet opening angle samples have a bin size of $\triangle z \sim 0.45$ and $\triangle \theta_j \sim 0.03$ radian, respectively, and each data point (filled circles) has been normalized by its corresponding distribution. The observed distributions have a mean redshift of approximately $\sim 1.5$ for the pre\Swift~sample and $z \sim 2$ for the~\Swift~sample, while the mean jet opening angle is approximately $\sim 0.1$ radian for the~\Swift-Ryan sample. These values motivated the adopted bin sizes. From the plots, the observed burst rate distributions increase steadily from low redshift, peaks near $z \sim 1.5$ and $z \sim 2$ for pre\Swift-Friedman and \Swift-Perley samples, respectively, and then decline toward a higher redshift. In contrast, the observed burst rate per jet opening angle distributions decreases rapidly up to $\theta_{j} \sim 0.2$ radian and remains approximately constant beyond that value.

\section{GRB MODEL OVERVIEW AND PARAMETER SELECTION METHOD} 
In this section, we briefly recap the LD07 framework including the burst-rate of GRB calculations and our fitting methodology. To calculate the differential burst-rate per redshift and the peak-flux distributions, we use LD07 Equation (16) and Equation (18), respectively, as

$$
\frac{d\dot{N}(> \fluxthres)}{d\Omega \ dz }  =  \frac{c g_0}{H_0
(2+s)} \frac{d^2_L(z) \ \dot{n}_{co}(z) }{(1+z)^3 \ \sqrt{\Omega_m
(1+z)^3 + \Omega_\Lambda}} \;
$$
\begin{equation}
\times \; \{ [1-\max(\hat\mu_j,\mu_{\rm jmin})]^{2+s} -
(1-\mu_{\rm jmax})^{2+s}\}\;, \label{eq1}
\end{equation}

$$
\frac{d\dot{N}(> \fluxthres)}{d\Omega  }  =
\frac{c g_0}{H_0
(2+s)} \int_0^{\zmax} dz\;\frac{d^2_L(z) \ \dot{n}_{co}(z)
}{(1+z)^3 \ \sqrt{\Omega_m (1+z)^3 + \Omega_\Lambda}} \;
$$
\begin{equation}
  \times \;\; \{ [1-\max(\hat\mu_j,\mu_{\rm jmin})]^{2+s} - (1-\mu_{\rm jmax})^{2+s}\}\;,
\label{eq2}
\end{equation}
and to calculate the differential burst-rate per jet opening-angle distribution, we use LD07 Equation (20) as
\begin{eqnarray}
\frac{d\dot{N}(> \fluxthres)}{d\Omega d\mu_j} & = & \frac{c}{H_0}
g(\mu_j)(1-\mu_j) \\
\nonumber
& \times & \; \int_0^{\zmax(\mu_j)}\frac{d^2_L(z) \
\dot{n}_{co}(z) \; dz }{(1+z)^3 \ \sqrt{\Omega_m (1+z)^3 +
\Omega_\Lambda}} \ , 
\label{eq3}
\end{eqnarray}
or 
\begin{eqnarray}
\frac{d\dot{N}(> \fluxthres)}{d\Omega d\theta} & = & -\sin \theta \, \frac{d\dot{N}(> \fluxthres)}{d\Omega d\mu_j}  \ , 
\label{eq4}
\end{eqnarray}
where $c$ and $\H0$ are the speed of light and the local Hubble constant, respectively. These equations describe (i) the differential directional GRB rate per redshift ($d\dot{N}(> \fluxthres)/d\Omega dz$), the differential directional GRB rate peak-flux distribution ($d\dot{N}(> \fluxthres)/d\Omega$), and (iii) the differential directional event reduction rate (due to the finite jet opening angle) for bursting sources per jet opening-angle ($d\dot{N}(> \fluxthres)/d\Omega d\theta$) with $\nu F_\nu$ spectral flux greater than $\fluxthres$ at the observed photon energy $\e$. These equations require the jet opening angle distribution $g(\mu_{\rm j})$, the comoving GRB rate density $\dot{n}_{co}(z)$, and the instrument's detector sensitivity $\fluxthres$. In this paper, we continue to use $\fluxthres \sim 10^{-8} $ and $\sim 10^{-7}$ erg cm$^{-2}$ s$^{-1}$ (see Table~\ref{tbl-2} and LD07) as the effective \Swift~and pre\Swift~detective flux thresholds, respectively. Since the form for the jet opening angle $g(\mu_{\rm j})$ is unknown, we also consider the function
$g(\mu_{\rm j}) = g_0 \ (1-\mu_{\rm j})^s \ H(\mu_{\rm j};\mu_{\rm j,min},\mu_{\rm j,max})$,
where $s$ is the jet opening angle power-law index; for a two-sided jet, $\mu_{\rm j,min} \geq 0$, and 
$g_0 = (1+s)/((1-\mu_{\rm j,min})^{1+s} - (1-\mu_{\rm j,max})^{1+s})$ 
is the distribution normalization (LD07), and $H(\mu; \mu_{\rm j}, 1)$ is the heaviside function such that $H(\mu; \mu_{\rm j}, 1) = 1$ when $\mu_{\rm j} \le \mu \le1$ (or when the angle $\theta$ of the observer with respect to the jet axis is within the opening angle of the jet), and $H(\mu; \mu_{\rm j}, 1) = 0$ otherwise. 

According to the observations from the pre\Swift~and \Swift~instruments, the minimum and maximum jet opening angles are about $\theta_{\rm j,min} \sim 0.042$ and $\theta_{\rm j,max} \sim 0.7$ rad, where $\mu_{j} = cos(\theta_j)$~\citep[e.g.][]{fb05,rya15,zha15}. This functional form $g(\mu_{\rm j})$ describes GRBs with small opening angles, $\theta_{\rm j} \ll 1$, so that such GRBs are potentially detectable from larger distances, for the same energy budget. By contrast, GRB jets with large opening angles are more frequent, but only detectable from comparatively small distances (e.g. LD07). In our model, the apparent divergence at $\mu_{j} = 1$ is controlled by imposing finite lower and upper limits on the jet opening angle. Specifically, the distribution is only evaluated over the physically allowed range $\theta_{\rm j,min} \leq \theta_j \leq \theta_{\rm j,max}$, so it never reaches the unphysical limit $\theta_{\rm j} \rightarrow 0$, or equivalently $\mu_{\rm j} \rightarrow 1$. Therefore, the model remains finite within the adopted jet opening angle range (see LD07 for details).

The possibility of the beaming angle being narrower at higher redshifts has also been suggested by others \citep[e.g.][]{lfr02,lu12,las14,las18a,las18b,laj19,lha19}.
These requirements dictate that the maximum redshift $z_{max}$ must satisfy the condition
$d^2_L(\zmax) \; \leq \; \Estarg/(4 {\rm \pi} (1 - \mu_{\rm j,min}) \; \Delta t_* \; \fluxthres \; \lambda_b)$, where $d_L(z) = \frac{c}{H_0}(1+z) \ \int^z_0 \frac{dz^\prime}{\sqrt{\Omega_m (1+z^\prime)^3 + \Omega_\Lambda}}$ is the luminosity distance, $\Delta t_*$ is the duration of the GRB in the stationary frame and we set it equal to $10 s$ as the average observed value based on BATSE GRBs at $z \sim 1$, $\Estarg$ is the mean beaming corrected $\gamma$-ray energy release and it is about $2 \times 10^{51}$ ergs~\citep[see][]{fb05}, and $\lambda_b$ is the bolometric correction to the peak measured $\nu F_\nu$ flux as described by Equations (1) and (2) in LD07 with $\lambda_b = (a^{-1} - b^{-1})$, where $a$ and $b$ are the broken power-law indices of the GRB spectrum. We refer the readers to section 2 of LD07 for more details of our model.
Finally, to close the system, the comoving GRB rate density is defined as
$\dot{n}_{co}(z) = \dot{n}_{co} \Sigma_{_{\rm SFR}}(z)$,
where $\dot{n}_{co}$ is the normalization constant and $\Sigma_{_{\rm SFR}}$ is the GRB formation rate as a function of redshift of the form
\begin{equation}
    \Sigma_{_{\rm SFR}}(z)=\left\{
                \begin{array}{ll}
                  a_0 (1+z)^{\eta_{_1}} \hskip+0.25in , 0 \leq z \leq z_1 \\
                  b_0 (1+z)^{\eta_{_2}} \hskip+0.25in , z_1 < z \leq z_2  \;, \\
                  c_0 (1+z)^{\eta_{_3}} \hskip+0.25in , z > z_2 
                \end{array}
              \right.  
\label{eq5}
\end{equation}
where $a_0 = 1$, $b_0 = a_0 (1+z_1)^{\eta_{_1}} / (1+z_1)^{\eta_{_2}}$, $c_0 = b_0 (1+z_2)^{\eta_{_2}} / (1+z_2)^{\eta_{_3}}$, and $\eta_{_1}, \eta_{_2}$, $\eta_{_3}$ are constants, and the redshift values $z_1$ and $z_2$ are equal to $0.5$ and $4.5$, respectively.  We select these break redshifts because they were shown in LM17 and LRM20 to provide consistent agreement with the observed distributions, and we retain the same values in the present work to examine how variations in $\H0$ affect the predicted differential burst-rate distributions. Table~\ref{tbl-1} lists the GRB formation-rate model parameters ($\eta_{_1}, \eta_{_2}, \eta_{_3}$) for the SFR9 and SFR11 model discussed in Section~2. For reference, SFR7 model, which represents the observed \citet[][]{hb06} cosmic star formation-rate history extended by \citet[][]{li08} is described by $\Sigma_{_{\rm SFR}}(z) = \frac{1+(\eta_{_2} z/\eta_{_1})}{1+(z/\eta_{_3})^{\eta_{_4})}}$ as a function of redshift $z$. The parameter values $\eta_{_1}, \eta_{_2}, \eta_{_3}$, and $\eta_{_4}$ for the SFR5, SFR6, SFR7 models are given in Table~\ref{tbl-1}. As reviewed in Section 2, LD07 showed that the SFR5 and SFR6 models reproduce the overall behavior of the cumulative redshift and jet opening-angle distributions of the early pre\Swift-Friedman and~\Swift-Jakobsson GRB samples.

The constant $\eta_{_1}, \eta_{_2}$, $\eta_{_3}$ from equation (\ref{eq5}), together with the power-law index $s$, are  varied until the calculated model distributions provide consistent agreement with the observed pre\Swift~and \Swift~redshift and jet opening angle distributions for a fixed set of the physical parameters ($H_0, a, b, \theta_{\rm j,min}, \theta_{\rm j,max}, \Estarg,$ and $\fluxthres$). If the model does not adequately reproduce the observed distributions, the physical parameter values are readjusted and the model distributions recalculate. These two steps are iterate until agreement is obtained between the predicted and observed distributions. This procedure allows us to identify parameter values that provide a consistent description of the observation.

As a final validation step, we apply the Kolmogorov-Smirnov (KS) test to determine whether the observed cumulative redshift and jet opening-angle distributions are statistically consistent with the model predictions, based on the KS ($D$)-statistic and the corresponding ($p$)-value. As discussed in Section 2, the two-sample KS test shows that the pre\Swift-Friedman and~\Swift-Perley redshift distributions are statistically consistent with being drawn from the same parent population. Furthermore, the one-sample KS tests yield p-values greater than $0.05$ for the adopted model predictions of the pre\Swift-Friedman redshift,~\Swift-Perley redshift, and~\Swift-Ryan jet opening-angle distributions, indicating no statistically significant disagreement between the model and the observations. Details of the fitting procedure are given in Section 3.2 of LD07.

In this paper, we use the GRB formation-rate models SFR7, SFR9, and SFR11 (see Table~\ref{tbl-1} and Figure~\ref{fig1}), following the framework developed by LMR20 and discussed in Section~2, to investigate how the predicted GRB distributions vary with $\H0$. Previous work demonstrated that, among the GRB formation-rate models considered, SFR11 model provides consistent agreement with the cumulative redshift and jet opening-angle distributions of the combined pre\Swift~and~\Swift~samples. In the present study, we impose the additional requirement that the model must simultaneously reproduce the differential burst-rate-per-redshift and differential burst-rate-per-opening-angle distributions. The preferred solution is therefore the parameter set that best describes both the cumulative and differential distributions.

\section{RESULTS AND DISCUSSION}
Similar to the approach in LD07, LM17, and LRM20, there are seven adjustable physical parameters in this model: the $\nu F_\nu$ spectral power-law indices $a$ and $b$, the power-law index $s$ of the jet opening-angle distribution, the range of the jet opening angles $\theta_{\rm j,min}$ and $\theta_{\rm j,max}$, the average absolute emitted gamma-ray energy $\Estarg$, and the detector threshold $\fluxthres$ excluding the Hubble constant $H_0$ and the constant values of $\eta_{_1}, \eta_{_2}$, and $\eta_{_3}$ in the GRB rate density. We adopt the values of $\eta_{_1}, \eta_{_2}$, and $\eta_{_3}$ as listed in Table~2, together with $s = -1.55$, and $\Estarg = 4.47 \times 10^{51}$ erg as in LRM20, since these choices provide consistent agreement with the observed redshift and jet opening angle distributions. The flux thresholds $\fluxthres$ are set equal to $10^{-8}$  and $10^{-7}$ erg cm$^{-2}$ s$^{-1}$ for \Swift~and pre\Swift~samples, respectively, and we set the observed jet opening angles to $\theta_{\rm j,min} = (0.04, 0.065)$ rad for \Swift~and pre\Swift~models, respectively, and $\theta_{\rm j,max} = 0.8$ rad according to the observed samples.  As noted earlier, we adopt a broken power-law $\nu F_\nu$ GRB SED with observed indices $\alpha=-0.9$ and $\beta = -2.18$, corresponding to $a =1$ and  $b = -0.5$, and $\lambda_b \approx 6.46$ as the bolometric correction factor. In this work, we hold these physical parameters fixed, as in LRM20, because they successfully reproduce both the observed redshift and jet opening-angle cumulative distributions. The corresponding parameter values ($\Estarg, \alpha, \beta, \theta_{\rm j,min}, \theta_{\rm j,max}, s$) are listed in Table~\ref{tbl-3} as Case-1 and Case-2. With these parameters fixed, we vary only the local Hubble constant, $\H0$, to examine how changes in $\H0$ affect the predicted differential burst-rate-per-redshift and jet opening-angle distributions of the~\Swift~and pre\Swift~samples.

\subsection{Model Parameters for the {\it Swift} and {\it preSwift} Redshift and Jet Opening Angle Samples}
For illustration, Figures~\ref{fig6}(a) and~\ref{fig6}(b) show the normalized differential event rate per unit redshift, $d{\dot{N}}/dz$, and the normalized differential event rate per unit jet opening-angle, $d{\dot{N}}/d\theta$, for $\H0 = 68, 72,$ and $75$ km s$^{-1}$ Mpc$^{-1}$, assuming \Swift-like sensitivity and GRB density rate model SFR11. From the model's perspective, increasing $\H0$ reduces both the luminosity distance $d_L$ and the comoving distance at a given redshift. This decreases in turn reduces the comoving volume element, leading to lower predicted GRB event rates across all redshift and jet opening angles. Consequently, the total number of detectable GRBs declines as $\H0$ increases, consistent with the trends shown in Figures~\ref{fig6}(a) and~\ref{fig6}(b). To compare the theoretical differential burst rate distribution with the observed differential, either as a function of redshift or jet opening angle, we normalize $d{\dot{N}}/dz d\Omega$ and $d{\dot{N}}/d\theta d\Omega$. This normalization removes the explicit directional dependence $\Omega$ and allows a direct comparison of the observed distributions. 

\begfig[t] \hskip-0.25in \epsscale{1.1} \plottwo{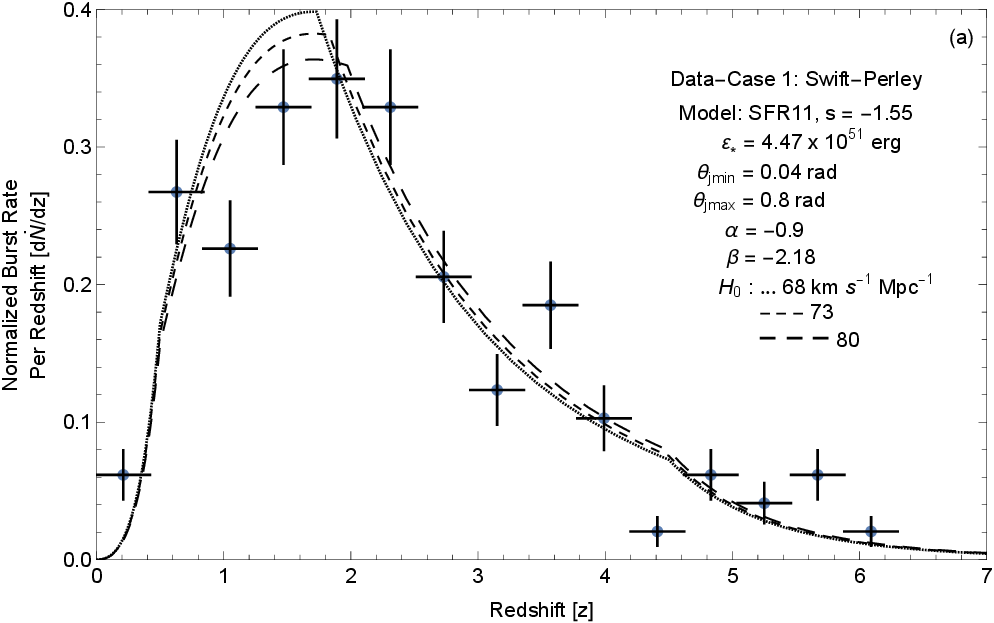}{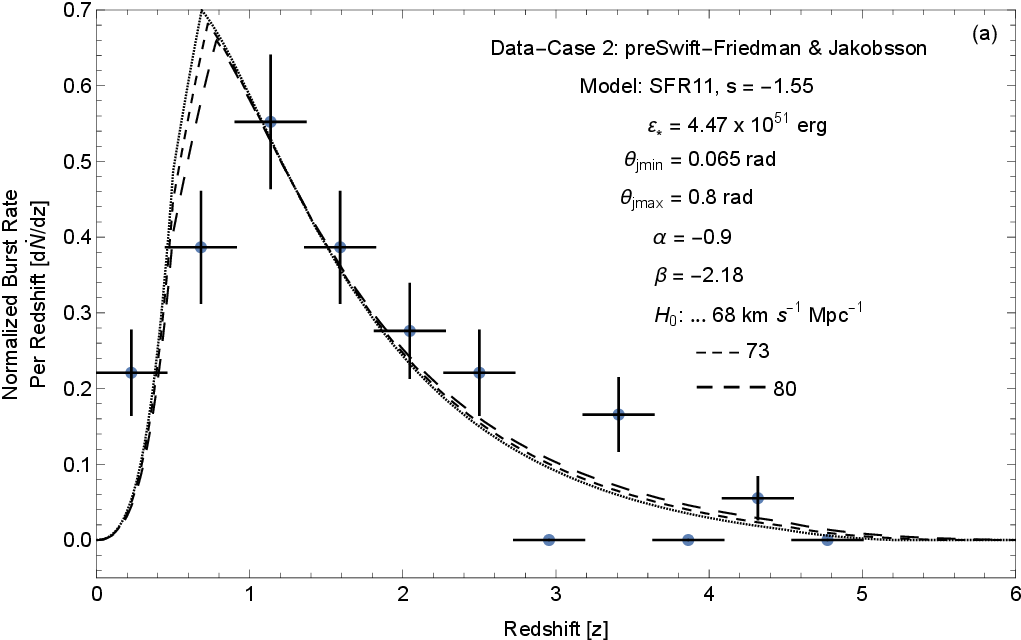}
\caption{\footnotesize ({\it a}) Observed normalized differential event rate per redshift distribution for the~\Swift–Perley sample, and ({\it b}) for the pre\Swift-Friedman \& Jakobsson sample. In both panels, the curves represent the model predictions based on the SFR11 GRB burst rate, with solid, short-dashed, and long-dashed lines corresponding to $H_0 = 68, 73, 80$ km s$^{-1}$ Mpc$^{-1}$, respectively.}
\label{fig7} 
\finfig
\begfig[t] \hskip-0.25in \epsscale{1.1} \plottwo{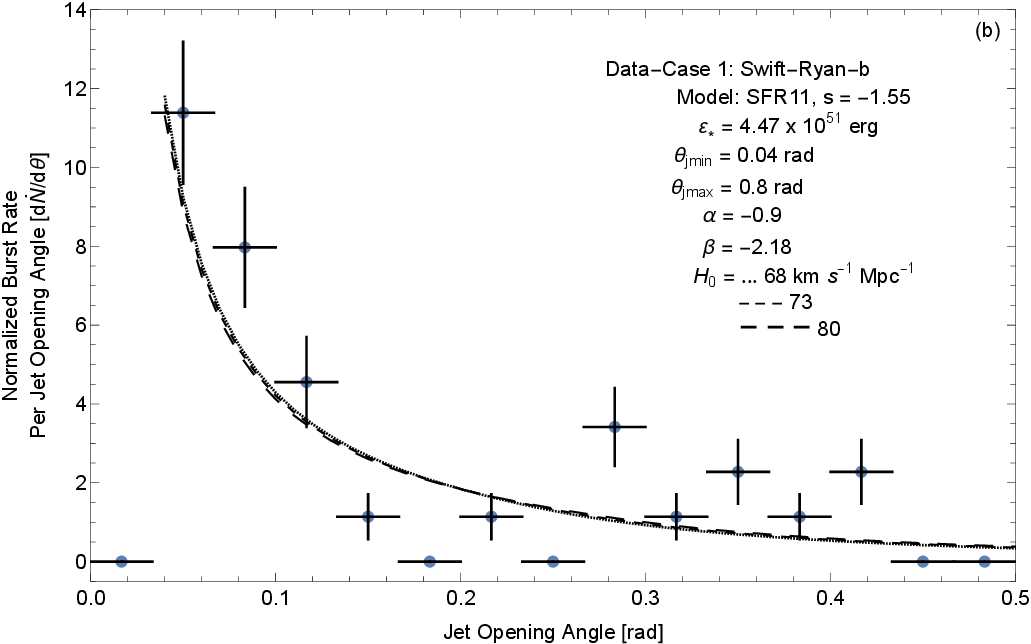}{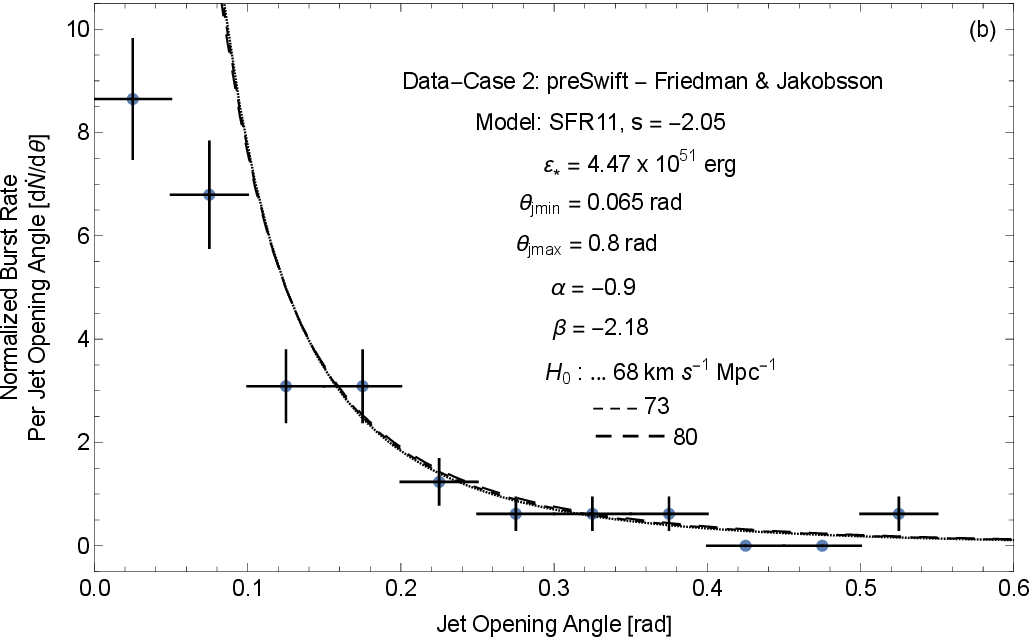}
\caption{\footnotesize ({\it a})  Observed normalized differential event rate per jet opening angle distribution for the~\Swift-Ryan-b sample, and ({\it b}) for the pre\Swift-Friedman \& Jakobsson sample. In both panels, the curves represent the model predictions based on the SFR11 GRB burst rate, with solid, short-dashed, and long-dashed lines corresponding to $H_0 = 68, 73, 80$ km s$^{-1}$ Mpc$^{-1}$, respectively.}
\label{fig8} 
\finfig
\begfig[t] \hskip-0.25in \epsscale{1.1} \plottwo{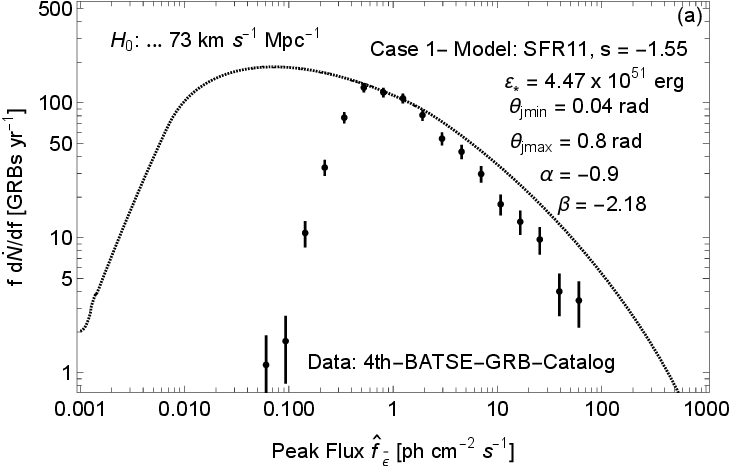}{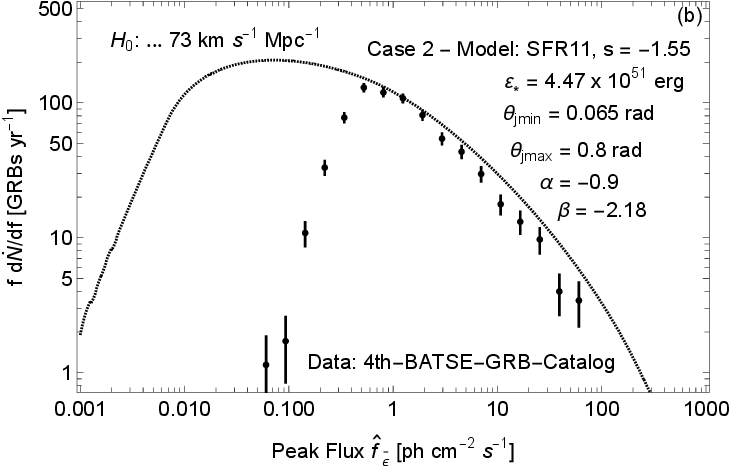}
\caption{\footnotesize ({\it a}) Differential {\bf peak-flux distribution} for the Case-1 model and ({\it b}) for Case-2 model. In both panels, the solid curves correspond to a Hubble constant of $73$ km s$^{-1}$ Mpc$^{-1}$. The filled circles represent the 1024 ms trigger-timescale data from the BATSE-4B catalog GRBs data, which includes 1292 bursts.}
\label{fig9} 
\finfig

Using the physical model parameters listed in Table 3, we apply the three Hubble constant values, $\H0 = 68, 73$, and $80$ km s$^{-1}$ Mpc$^{-1}$ to each of the Case-1 through Case-4 parameter sets. Case-1 (for the Swift-Perley and Swift-Ryan samples) and Case-2 (the preSwift-Friedman sample) are the adopted values from LRM20, Figures~\ref{fig7}(a)–\ref{fig7}(b) and Figures~\ref{fig8}(a)–\ref{fig8}(b) show the predicted differential GRB burst-rate distributions assuming the SFR11 GRB formation-rate model, while their associated cumulative redshift and jet opening angle distributions are shown in Figure~\ref{fig2}. The solid, short-dashed, and long-dashed curves in Figures~\ref{fig7} and~\ref{fig8} correspond to Hubble constant values of $H_0 = 68, 73$, and $80$ km s$^{-1}$ Mpc$^{-1}$, respectively. For both the~\Swift~and pre\Swift~samples, the differential burst-rate distributions as a function of jet opening angle exhibit little to no dependence on the adopted value of $H_0$ (see Figures~\ref{fig8}(a)-\ref{fig8}(b)). In contrast, the differential redshift distributions show noticeable sensitivity to $H_0$ in the redshift range $1 \lesssim z \lesssim 2$ (see Figure~\ref{fig7}(a)) for the~\Swift~sample. In particular, the~\Swift~differential redshift distribution appears to favor a higher value of the Hubble constant, $H_0 \gtrsim 80$ km s$^{-1}$ Mpc$^{-1}$, as seen in Figure~\ref{fig7}(a); unless, the~\Swift~sample exhibits a deficit of GRBs in the redshift interval $1 \lesssim z \lesssim 2$. In addition, the pre\Swift~differential jet opening angle distribution is slightly shifted toward lower angles (see Figure~\ref{fig8}(b)). These features are not evident in the corresponding cumulative redshift or jet opening angle distributions (see Figures 2(a) and 2(b)), highlighting the importance of using the differential burst rate distributions in redshift and opening angle to reveal such effects. It is important to note that the zero-value data points in both distributions indicate bins in which no bursts were detected, rather than measured burst rates (see discussion near the end of Section~3). Therefore, the model is not expected to fit these points. For example, in Figure 8(a), the predicted differential jet opening-angle distribution is consistent with the overall trend of the observed distribution while not reproducing the zero-value point near $\theta_j \approx 0.02$ rad, which simply reflects the absence of detected bursts in that bin. In addition, the differential jet opening-angle distributions (see Figures~\ref{fig8}(a)-\ref{fig8}(b)) suggest that a slightly larger lower bound on the jet opening angle is required for the~\Swift~sample, while a smaller lower bound would provide better agreement with the pre\Swift~sample at lower jet opening angles. Moreover, the~\Swift~and pre\Swift~differential redshift distributions shown in Figures~\ref{fig7}(a) and~\ref{fig7}(b) further suggest that a lower GRB formation rate (similar to SFR9) below $z \approx 1.5$ may provide a better match to the data.

In Figures~\ref{fig9}(a)–\ref{fig9}(b), we plot the differential peak-flux distributions computed using the model parameters for Case-1 and Case-2 against the BATSE 4B-Catalog GRBs data. These curves did not show any sensitivity to the different values of the Hubble constant, hence, we only show for the case $\H0 = 73$ km s$^{-1}$ Mpc$^{-1}$. The primary difference between these two cases is the adopted lower limit of the jet opening angle which we have already discussed ($\theta_{\rm j,min}=0.04$ rad for Case-1 and $\theta_{\rm j,min}=0.065$ rad for Case-2). Both model slightly overpredict the number of high-flux bursts, indicating that beaming-corrected energy release in the model is likely too large. This result further suggests that better agreement with the observations could be obtained by adopting a minimum jet opening angle larger than $0.04$ radian for Case-1 and smaller than $0.065$ radian for Case-2, or by adopting a slightly more negative value of the jet opening-angle distribution power-law index $s$, which would shift the burst-rate distribution toward larger jet opening angles. From Figure~\ref{fig9}(a) and (b), below a photon number threshold of $0.3$ photons cm$^2$ s$^{-1}$ in the $50-300$ keV band, the observed number of GRBs falls rapidly due to the sharp decline in the BATSE trigger efficiency at these photon fluxes (see Paciesas et al. 1999). However, the peak-flux distribution of the Swift GRBs will extend to much lower values, $0.0625$ photons cm$^2$ s$^{-1}$, allowing us to use the model to predict the peak-flux distribution and compare it with the~\Swift~data, as discuss below in Figures 13 and 14 for model parameters in Case-3 and Case-4, respectively. We note, however, that the Swift triggering criteria are more complicated than a simple rate trigger, particularly near threshold \citep[][see also LD07 for more details]{ban06}.

%
\begfig[t] \hskip-0.25in \epsscale{1.1} \plottwo{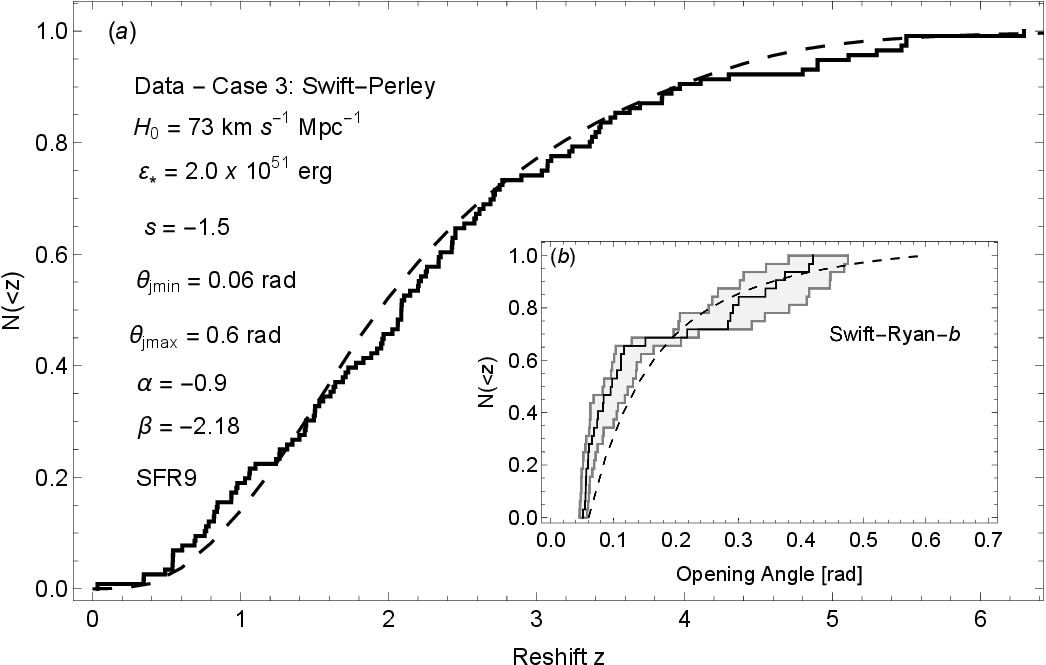}{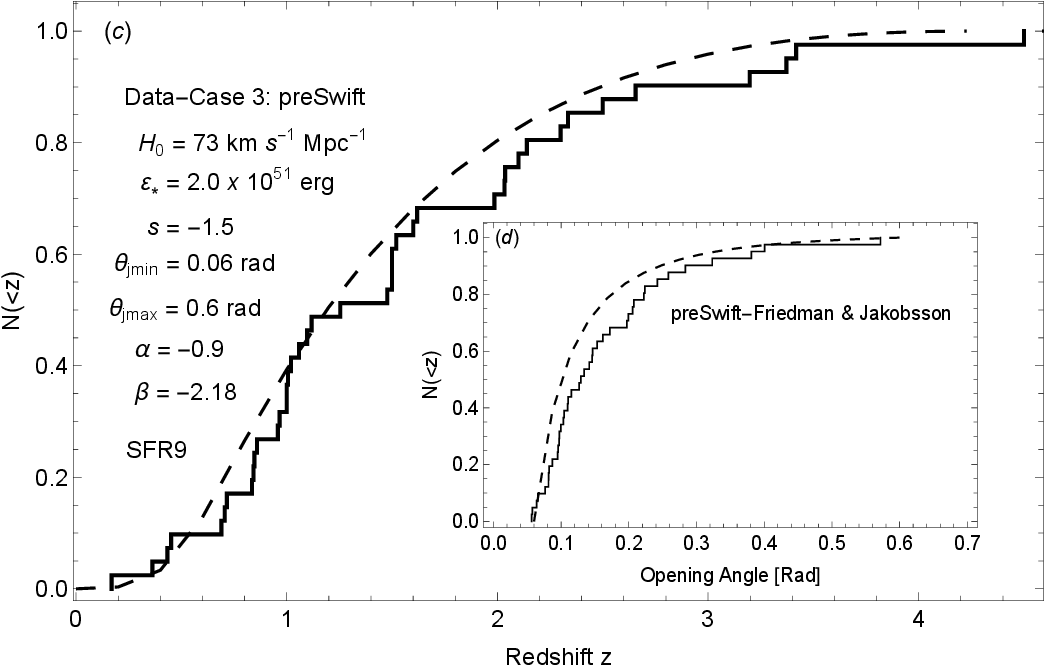}
\caption{\footnotesize Panel ({\it a}) and inset ({\it b}) show the~\Swift~redshift and jet-opening-angle cumulative distributions, while panel ({\it c}) and inset ({\it d}) show the corresponding pre-\Swift~distributions for the Case-3 model. The solid curves represent the observed~\Swift~(Perley; Ryan) and pre\Swift~(Friedman; Jakobsson) samples, with the shaded region in inset ({\it b}) indicating the~\Swift-Ryan uncertainties. The long-dashed curves in all panels show the model predictions assuming SFR9 model.}
\label{fig10} 
\finfig
\begfig[t] \hskip-0.25in \epsscale{1.1} \plottwo{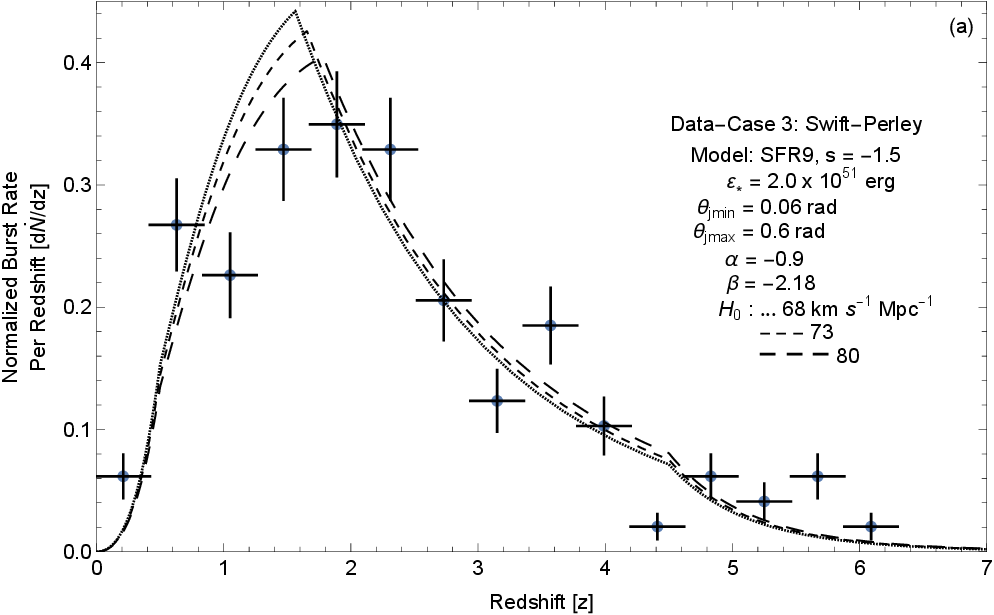}{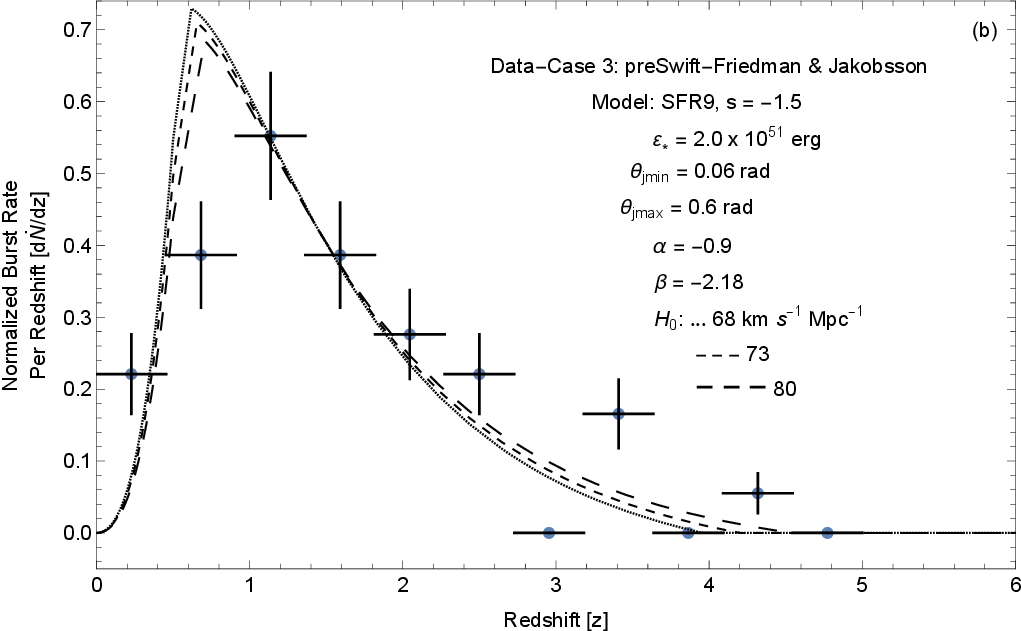}
\caption{\footnotesize ({\it a}) Normalized burst-rate per redshift distribution for the~\Swift-Perley sample and ({\it b}) the pre\Swift-Friedman \& Jakobsson sample. The long-dashed curves in both panels show the model predictions using Case-3 parameter values assuming SFR9 model.}
\label{fig11} 
\finfig
\begfig[t] \hskip-0.25in \epsscale{1.1} \plottwo{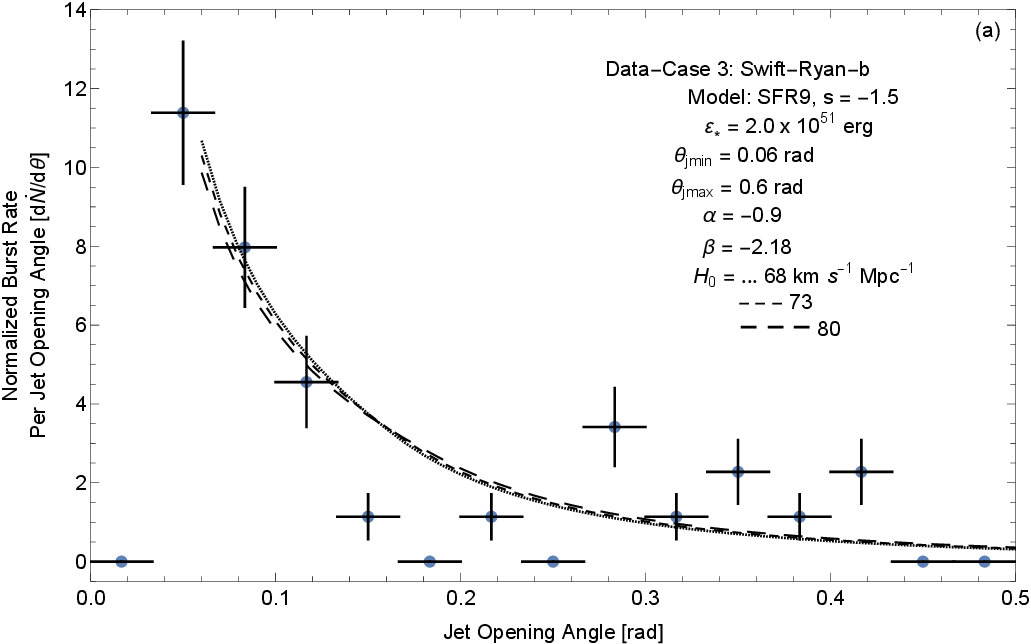}{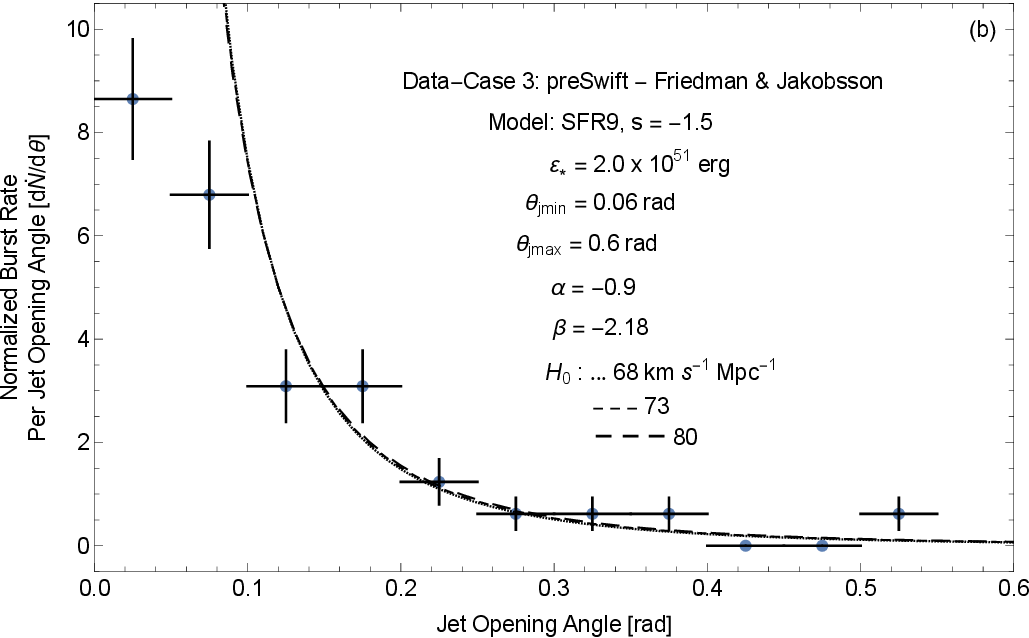}
\caption{\footnotesize ({\it a}) Normalized burst-rate per jet opening angle distribution for the~\Swift-Ryan-b sample and ({\it b}) the pre\Swift-Friedman \& Jakobsson sample. The long-dashed curves in both panels show the model predictions using Case-3 parameter values assuming SFR9 model.}
\label{fig12} 
\finfig
%
\begfig[t] \hskip-0.25in \epsscale{0.8} \plotone{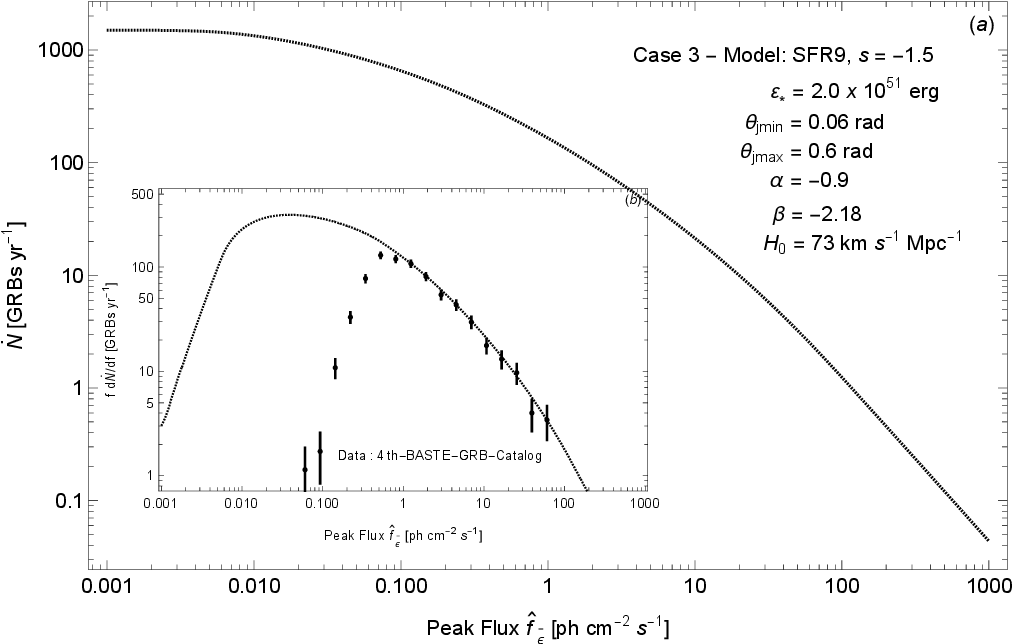}
\caption{\footnotesize ({\it a}) Integral size distributions and ({\it b}) the differential size distributions assuming SFR9 Case-3 model. The filled circles curve represents the 1024 ms trigger-timescale data from the 4B- Catalog, the same as in Figures~\ref{fig9}a and~\ref{fig9}b.}
\label{fig13} 
\finfig
Figures~\ref{fig10}(a)-\ref{fig10}(b), \ref{fig11}(a)-\ref{fig11}(b), and~\ref{fig12}(a)-\ref{fig12}(b) are analogous to Figures~\ref{fig2}(a)-~\ref{fig2}(b) and Figures~\ref{fig7}(a)-\ref{fig8}(b), but are computed using the GRB formation rate model SFR9 with Hubble constants of $H_0 = 68, 73, 80$ km s$^{-1}$ Mpc$^{-1}$ using Case-3 parameter values; these results also in favoring of a larger Hubble value. In Case-3, we adopt $s=-1.5, \theta_{\rm j,min} = 0.06$ rad, $\theta_{\rm j,max} = 0.6$ rad, and the reduce the beaming-corrected energy release to $\Estarg = 2.0 \times 10^{51}$ erg reflecting the analysis we just discuss above, and everything else remains the same as in Case-1 and Case-2 (see Table~\ref{tbl-3}). Similar to Figures~\ref{fig7} and~\ref{fig8}, the resulting distributions are largely insensitive to the adopted value of $H_0$ everywhere, except in the region of redshift $1-2$ for the~\Swift~sample of the redshift differential distribution (see Figures~\ref{fig11} and~\ref{fig12}). The cumulative and differential distributions obtained using the SFR9 model show overall good agreement with the observed data for both the~\Swift~and pre\Swift~samples (see Figures~\ref{fig10}-\ref{fig11}). The KS one-sample tests of the cumulative distributions yield p-values greater than $0.05$ for the adopted model predictions of the pre\Swift-Friedman redshift and jet opening angle,~\Swift-Perley redshift, and~\Swift-Ryan jet opening-angle distributions, indicating no statistically significant disagreement between the model and the observations. Most importantly, both the~\Swift~and pre\Swift~samples are described using the same set of model parameters (see Case-3 from Table~3), in contrast to previous studies where different lower limits on the jet opening angle were required to fit the two samples separately (e.g., see Figures \ref{fig7}(a) through \ref{fig9}(b) and Case-1 and Case-2 from Table~3). In addition, the differential peak-flux distribution shown in Figure~\ref{fig13}(b) provides an excellent agreement to the 4th BASTE GRB-catalog above a photon number threshold of $0.3$ photons cm$^2$ s$^{-1}$ in the $50-300$ keV band, improving upon the results obtained with the SFR11 model using the Case-1 and Case-2 parameters (see Figures~\ref{fig9}a and \ref{fig9}b). As we have discussed earlier in Figures 9(a) and 9(b), the observed number of GRBs falls rapidly below $0.3$ photons cm$^2$ s$^{-1}$ due to the sharp decline in the BATSE trigger efficiency at these photon fluxes. However, the peak-flux distribution of the Swift GRBs will extend to much lower values, $0.0625$ photons cm$^2$ s$^{-1}$, and we can use our model to predict the peak-flux distribution for comparison with the~\Swift~data as discuss below.

Figures~\ref{fig14}-\ref{fig16} and Figures~\ref{fig17}-\ref{fig19} are analogous to Figures~\ref{fig10}-\ref{fig12}, but are computed using the GRB formation rate models SFR11 and SFR7, respectively. We apply the same set of model parameters (Case-3) to both SFR11 and SFR7 to see how sensitive are these models to the observed differential distributions. While the differential redshift and jet opening-angle distributions yield results similar to those obtained with the SFR9 model, differences emerge in the cumulative redshift distribution for the~\Swift~sample. In particular, the results indicate that a substantially higher burst-rate density at high redshift is required, especially for the SFR7 model (see Figure~\ref{fig17}a). Figures \ref{fig20}(a) and \ref{fig20}(b) compare the predicted BATSE 4B peak-flux distributions for the SFR11 and SFR7 models, calculated using the Case-3 parameter values, with the observed data. The excellent agreement between the models and the BATSE 4B observations above a photon number threshold $0.3$ photons cm$^2$ s$^{-1}$ provides an independent validation of the adopted parameter set and suggests that the physical parameters are reasonably well constrained in comparison with model values Case-1 and Case-2, while weakly dependent on the GRB formation-rate model.

The above comparisons show an apparent discrepancy between the predicted and observed~\Swift~differential burst-rate per redshift distribution near $z \sim 1-2$, whereas a comparable discrepancy is not apparent in the pre\Swift~sample. Although the predicted~\Swift~differential redshift distribution shows sensitivity to variations in $\H0$ in this interval, this behavior should not be attributed uniquely to $\H0$. The shape of the differential redshift distribution also depends on the adopted GRB model and physical parameters, including the jet-opening-angle distribution index $s$, or the mean absolute emitted gamma-ray energy $\Estarg$. For example, Figures~\ref{fig11}(a),~\ref{fig15}(a), and~\ref{fig18}(a) use the same values of $s$ and $\Estarg$ but different GRB formation-rate models (SFR9, SFR11, and SFR7). Among these, SFR7 model shows a modest improvement in reproducing the observed differential observed burst-rate distribution near $z \sim 1-2$, suggesting that the assumed GRB formation-rate history may contribute to the discrepancy. 

The effects of $s$ and $\Estarg$ on the distributions can be seen by comparing Figures~\ref{fig7}(a) and~\ref{fig15}(a), which adopt the same GRB formation-rate model (SFR11) but different parameter values. A smaller value of $\Estarg$ shifts the differential burst-rate-per-redshift distribution toward lower redshifts, increasing the discrepancy between the model and the observed data near $z \sim 1-2$. At the same time, it shifts the differential burst-rate-per-opening-angle distribution toward larger jet opening angles. Similarly, a less negative value of $s$, corresponding to a reduced fraction of bursts with small jet opening angles (see Figure 3 of LD07), decreases the number of bursts observable at high redshift. This causes the differential burst-rate-per-redshift distribution to shift slightly toward lower redshifts (see Figures~\ref{fig7}(a) and~\ref{fig15}(a)), and the differential burst-rate-per-opening-angle distribution to shift toward larger jet opening angles (see Figures~\ref{fig8}(a) and~\ref{fig16}(a)). These comparisons suggest that the discrepancy near $z \sim 1-2$ may be more closely related to the adopted GRB formation-rate history, jet-population parameters, or the mean absolute emitted gamma-ray energy than to the value of the Hubble constant $\H0$. 

Figures~\ref{fig21}-\ref{fig24} are analogous to Figures~\ref{fig10}-\ref{fig13}, but for Case-4 parameter values with the same GRB formation-rate model SFR9. In Case-4, we adopt $s=-1.55$, $\theta_{\rm j,min} = 0.055$ rad, $\theta_{\rm j,max} = 0.6$ rad, and the beaming-corrected energy release $\Estarg = 2.51 \times 10^{51}$ erg reflecting the analysis we just discuss above. Both the predicted cumulative and differential distributions are consistent with the observed pre\Swift~and~\Swift~samples for the Hubble constants of $H_0 = 68, 73, 80$ km s$^{-1}$ Mpc$^{-1}$. The KS one-sample tests of the cumulative distributions yield p-values greater than $0.05$, indicating no statistically significant disagreement between the model and the observations. Most importantly, with Case-4 parameters, the discrepancy previously seen near $z\sim1-2$ is no longer apparent, as the predicted differential redshift distributions for all three values of $\H0$ are similar to the observed~\Swift~distribution in the redshift range. This suggests that the earlier discrepancy is not a robust feature of the data, but depends on the adopted GRB model parameters. These results also suggest that the differential distribution is only weakly sensitive to the variations in the Hubble constant. Figure \ref{fig24} also shows agreement between the model and the BATSE 4B observations above a photon number threshold $0.3$ photons cm$^2$ s$^{-1}$. 

\FloatBarrier
\begfig[t] \hskip-0.25in \epsscale{1.1} \plottwo{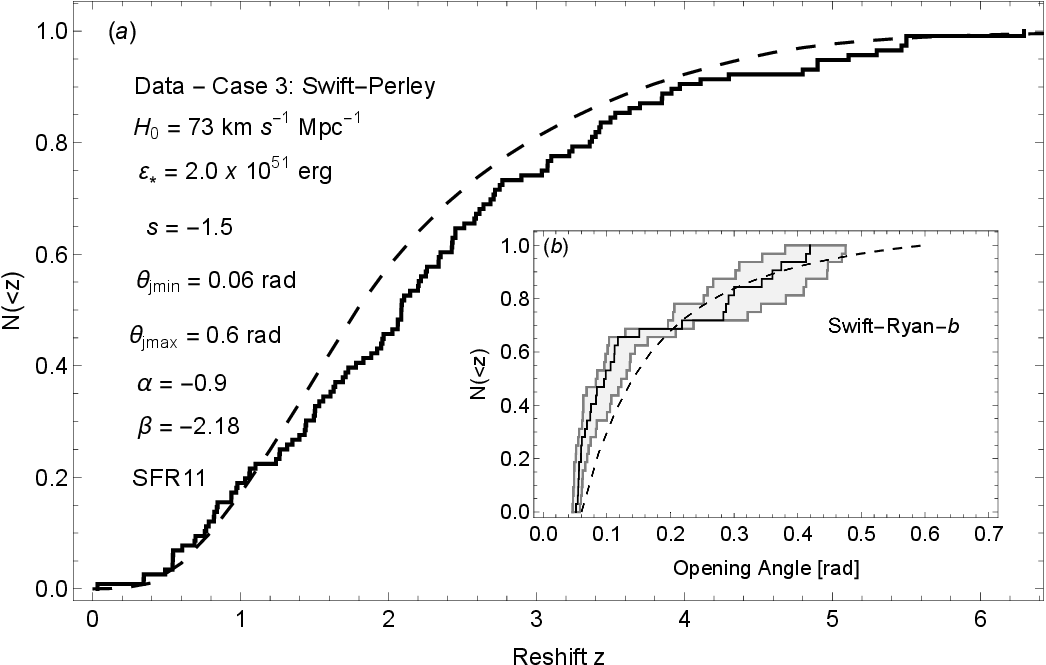}{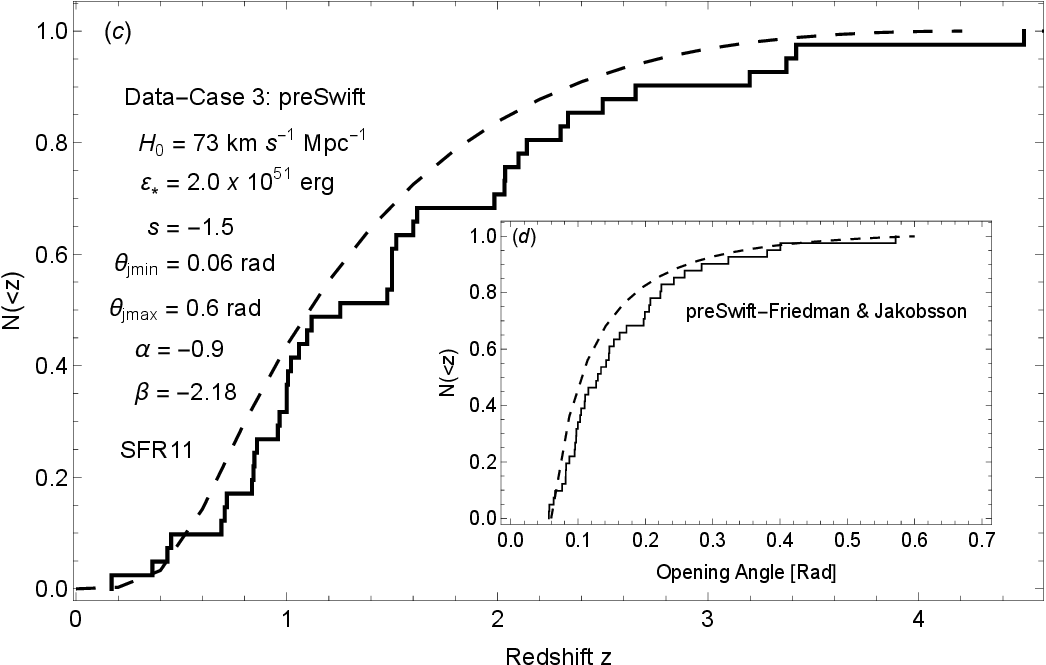}
\caption{\footnotesize Panels ({\it a}) and inset ({\it b}) show the~\Swift~redshift and jet opening angle cumulative distributions, while panels ({\it c}) and inset ({\it d}) show the corresponding pre\Swift~distributions using Case-3 parameter values, analogous to Figures 10({\it a}) and 10({\it b}) but for the SFR11 model.}
\label{fig14} 
\finfig
\begfig[t] \hskip-0.25in \epsscale{1.1} \plottwo{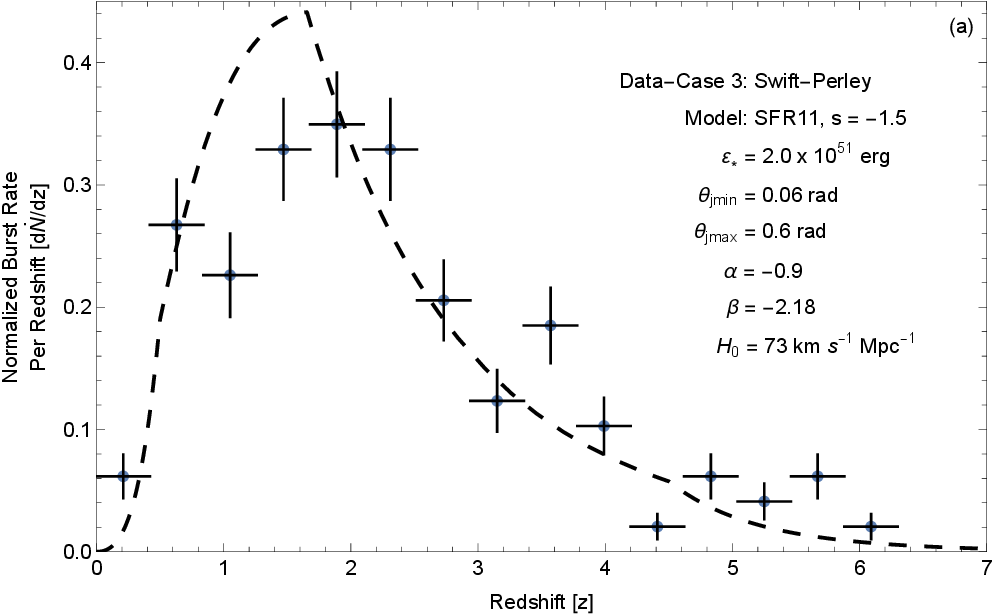}{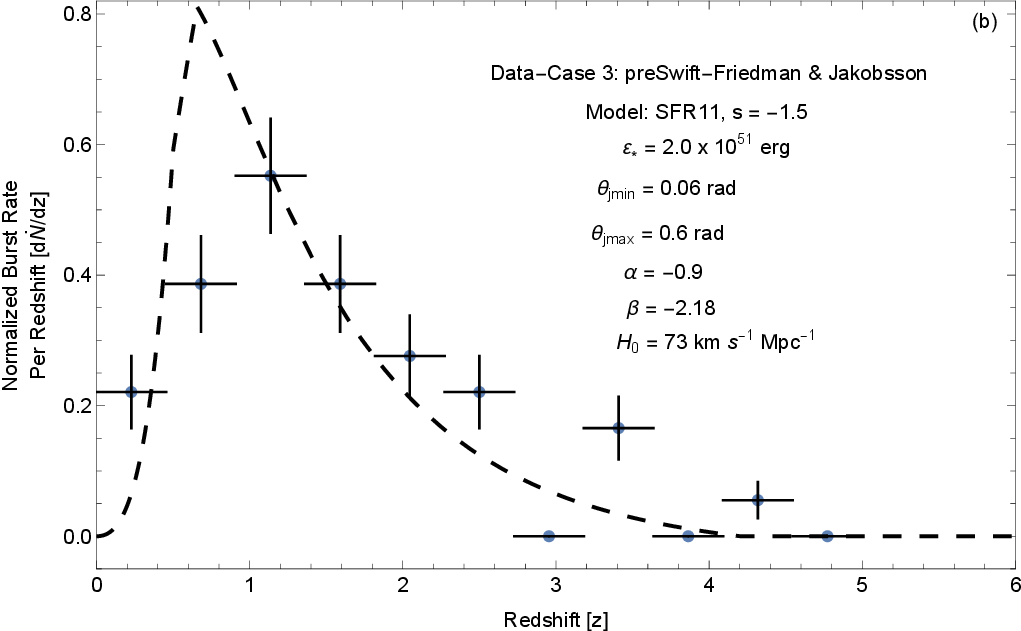}
\caption{\footnotesize Panels ({\it a}) and ({\it b}) show the normalized burst rate per redshift distribution for the~\Swift~and pre\Swift~samples using Case-3 parameter values, analogous to Figures 11({\it a}) and 11({\it b}) but for the SFR11 model.}
\label{fig15} 
\finfig
\begfig[t] \hskip-0.25in \epsscale{1.1} \plottwo{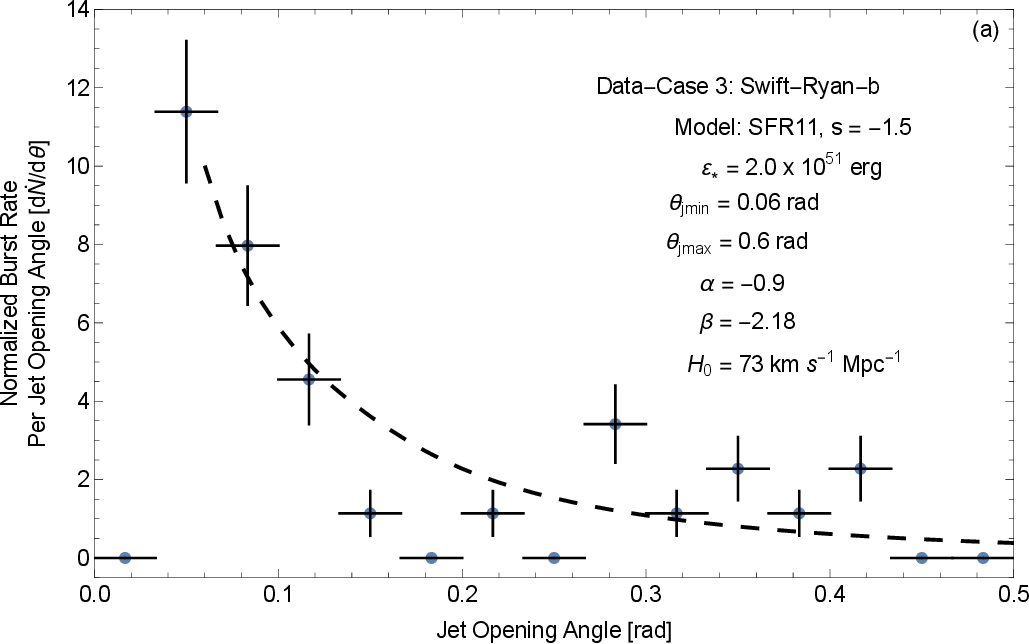}{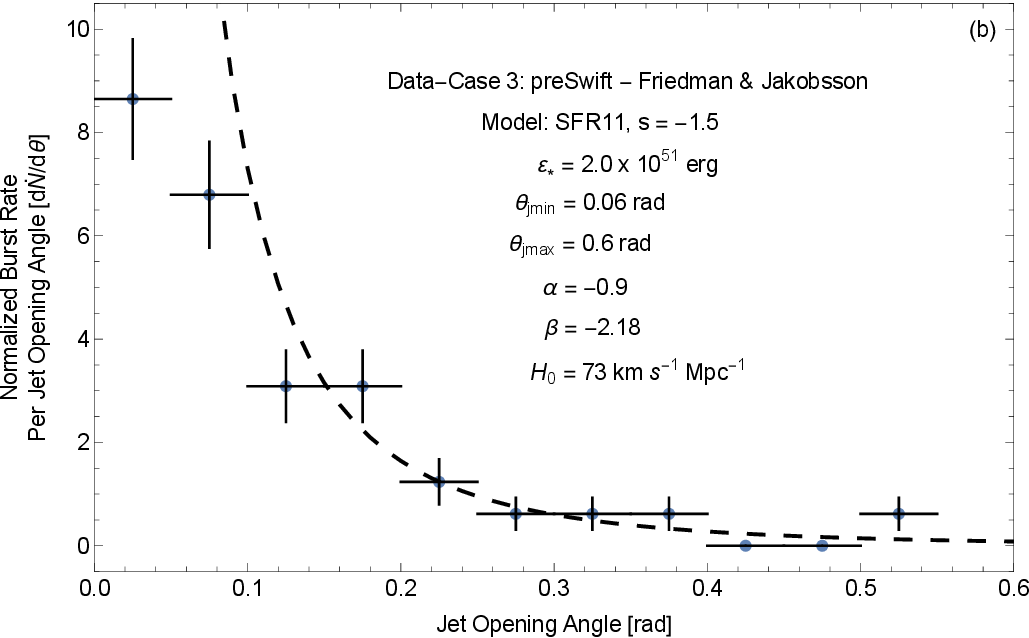}
\caption{\footnotesize ({\it a}) Panels ({\it a}) and ({\it b}) show the normalized burst rate per jet opening angle distribution for the~\Swift~and pre\Swift~samples using Case-3 parameter values, analogous to Figures 12({\it a}) and 12({\it b}) but for the SFR11 model.}
\label{fig16} 
\finfig
%

It is also important to note that the SFR7 model represents the observed cosmic star-formation rate density (see Table~\ref{tbl-1}), which we adopt to describe the GRB formation rate. Using this model, we obtain good agreement with the observed differential burst-rate distributions in redshift and jet opening angle for both the~\Swift~and pre\Swift~samples. The SFR9 and SFR11 models also provide consistent agreement with the observations, as discussed above. These results indicate that differential burst-rate distributions in redshift, jet opening angle, or peak-flux alone provide relatively weak constraints on the evolution of the long-GRB comoving rate density and may lead to biased inferences when considered in isolation. More robust constraints are obtained when the differential distributions are analyzed jointly with the corresponding cumulative redshift and jet opening-angle distributions, as well as the differential peak-flux distribution, as demonstrated for the SFR9 in Case-3 and Case-4. Finally, we emphasize that Case-4 parameter values are consistent within the observed physical constraints, including the GRB spectral energy distribution power-law indices $\alpha$ and $\beta$, the range of the jet opening angles $\theta_{\rm j,min}$ and $\theta_{\rm j,max}$, the average absolute gamma-ray energy release $\Estarg$, and the detector threshold $\fluxthres$ for both the~\Swift~and pre\Swift~instruments. Within this physically and observationally consistent parameter set, the Hubble constant in the range $H_0 = 68 - 80$ km s$^{-1}$ Mpc$^{-1}$ produce cumulative, differential, and peak-flux distributions that are consistent with the observations considered in this work.

\subsection{Estimating the Beaming Factor and the Luminosity Function}
Using the physically and observationally consistent parameter sets adopted in Case-3 and Case-4, we estimate the average beaming correction factor $<f_b>^{-1}$, where the beaming factor is averaged over the opening-angle distribution. This average is computed following Equation(21) of Le \& Dermer, which is given by $<f_b> = \left(\frac{1+s}{2+s}\right) \left[\frac{(1-\mu_{jmin})^{2+s} - (1-\mu_{jmax})^{2+s}}{(1-\mu_{jmin})^{1+s} - (1-\mu_{jmax})^{1+s}} \right]$. Using the parameter values adopted in Case-3(Case-4), we obtain an average beaming correction factor of $<f_b>^{-1} \approx 60(67)$, respectively. These values are fully consistent with the beaming factors inferred in previous studies \citep[e.g.,][]{bfk03, ggl04,gpw05,ggs12}, indicating that the jet geometry adopted in our model is in good agreement with independent constraints from GRB population studies. Following previous studies of jet-induced luminosity by Le \& Dermer, the luminosity function infers from the jet opening angle distribution is expressed as $dN/dL_* \propto L^{-(s+2)}_*$, where $L_*$ is the GRB apparent isotropic luminosity. We note a minor typographical error in Le \& Dermer (2007), where the induced luminosity function is stated as $dN/dL_* \propto L_*^{-3.25}$ rather than $ \propto L_*^{-0.75}$ for $s \sim -1.25$. The adopted jet opening-angle power-law indices $s = -1.5$ and $-1.55$ correspond to apparent isotropic luminosity functions $dN/dL_*  \propto L_*^{-0.5}$ and $L_*^{-0.45}$, respectively, within the uniform jet framework.

Our uniform-jet model induces a much shallower isotropic-equivalent luminosity function than the empirical broken power-law form inferred by \citet{wp10}, indicating that luminosity diversity in our model is primarily geometric rather than intrinsic. Moreover, \citet{pgs15} demonstrated that the apparent isotropic-equivalent luminosity function of long GRBs is strongly shaped by jet geometry and viewing angle effects. They showed that a narrow intrinsic energy distribution combined with realistic jet structures can reproduce the observed~\Swift~luminosity and redshift distributions without invoking intrinsic luminosity evolution. Our results are fully consistent with this interpretation and further show that a geometry-induced luminosity function derived from the fitted jet opening-angle distribution provides a unified description of~\Swift~and pre\Swift~observations.

\subsection{GRB Detection Rates and Model Validation}
To assess whether the physical parameters adopted in Case-3 and Case-4 and the GRB formation-rate model SFR9 provide an appropriate description of the~\Swift~observations, we examine the predicted GRB detection rates. Using the Case-3 (Case-4) parameters, our model predicts that~\Swift~can detect LGRBs out to a maximum redshift of $z \sim 11(14)$, and that approximately 10\% of~Swift~LGRBs should occur at $z \,>\, 4$, consistent with the data shown in Figures~\ref{fig10}(a) and~\ref{fig21}(a). The model further predicts that about 5\% of LGRBs should be detected at redshifts $z \,>\,  4.5$. If fewer than 5\% of LGRBs are observed above $z \,>\,  4.5$ by the time $\sim 100$~\Swift~LGRBs with measured redshifts have been accumulated, this would support the conclusion that the LGRB formation rate follows the SFR9 model, which is similar to the \citet{hb06} star formation history. In addition, the model predicts that fewer than 15\% of LGRBs should occur at low redshift ($z \,\leq\, 1$) under the SFR9 scenario. Examining the Swift GRB samples between 2013 and 2015, we find that fewer than 5\% of LGRBs per year were detected at $z \,>\,5$, approximately 10\% at $z \,>\,4$, and only about 6\% at $z \,<\, 1$. These observed fractions are in good agreement with our model predictions, indicating that the SFR9 model provides a plausible representation of the LGRB comoving rate density.

We further compare our model peak-flux distribution with BATSE 4B Catalog size distribution. BATSE 4B Catalog contains a total of 1292 GRBs, including both short and long duration events. Of these, 872 are identified as long-duration GRBs \citep[e.g.,][]{pmp99,ld09}. The BATSE detection rate corresponds to approximately 550 GRBs per year full-sky brighter than $0.3$ photons cm $^{-2}$ s$^{-1}$ in the 50-300 keV energy band \citep[][]{ban02}. Scaling by this fraction of LGRBs in the catalog (872/1292), this implies a BATSE full-sky detection rate approximately 371 LGRBs per year. Figures~\ref{fig13}(a) and~\ref{fig14}(a) show the model integral peak-flux distribution of LGRBs predicted by our best model, SFR9. The distributions are normalized to the observed BATSE LGRBs rate of 371 bursts over $4 \pi$ sr above a peak flux of 0.3 photons cm$^{2}$ s$^{-1}$ in the $50 - 300$ keV band for the $\triangle t = 1.024$ s trigger timescale. From the model peak-flux distribution, we find that approximately 300 LGRBs per year should be detected by a BATSE-like detector over the full sky above an energy-flux threshold of $\sim 10^{-7}$ ergs cm$^{-2}$ s$^{-1}$, or photon number threshold of $\gtrsim 0.633$ ph cm$^{-2}$ s$^{-1}$ assuming SFR9 rate model. In addition, our {\bf analysis} indicate that roughly 800 LGRBs occur per year over $4 \pi$ sr with peak fluxes exceeding $10^{-8}$ ergs cm$^{-2}$ s$^{-1}$, or $\gtrsim 0.0633$ ph cm$^{-2}$ s$^{-1}$ for both Case-3 and Case-4. Given that the field of view of the~\Swift~BAT instrument is approximately 1.4 sr \citep[][]{geh04}, our model predicts that~\Swift~should detect about 90 LGRBs per year under the SFR9 scenario. Observationally,~\Swift~detects roughly 135 GRBs per year in total; applying the same LGRB fraction (872/1292) yields an observed rate of approximately 90 LGRBs per year. This agreement provides additional support for the SFR9 model as a plausible description of the LGRB formation rate.

\begfig[t] \hskip-0.25in \epsscale{1.1} \plottwo{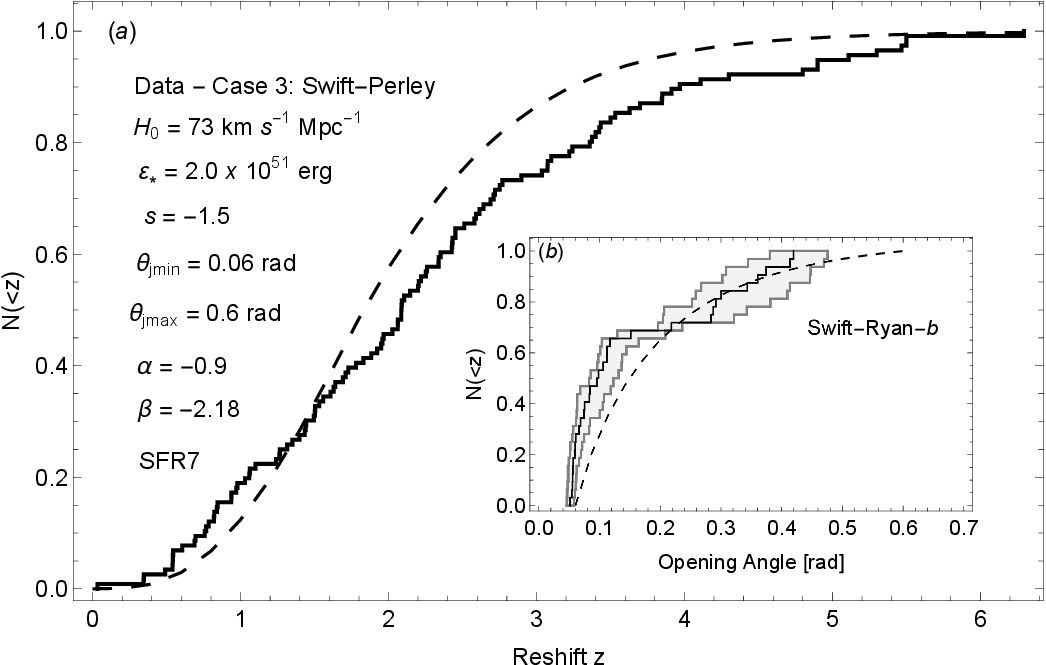}{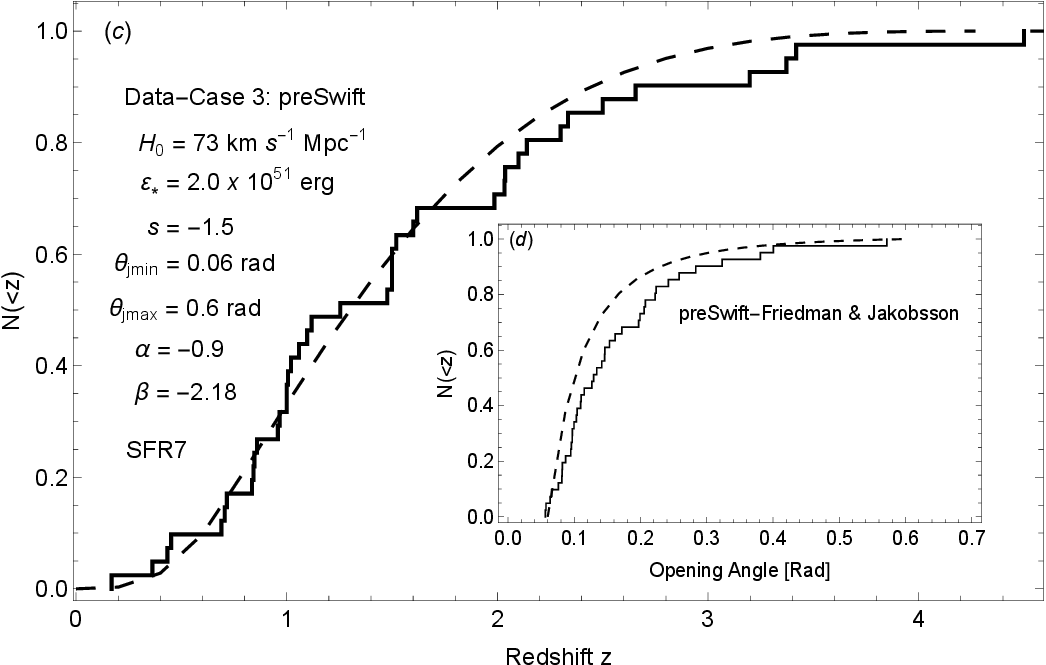}
\caption{\footnotesize Panels ({\it a}) and inset ({\it b}) show the~\Swift~redshift and jet opening angle cumulative distributions, while panels ({\it c}) and inset ({\it d}) show the corresponding pre\Swift~distributions using Case-3 parameter values, analogous to Figures 10({\it a}) and 10({\it b}) but for the SFR7 model.}
\label{fig17} 
\finfig
\begfig[t] \hskip-0.25in \epsscale{1.1} \plottwo{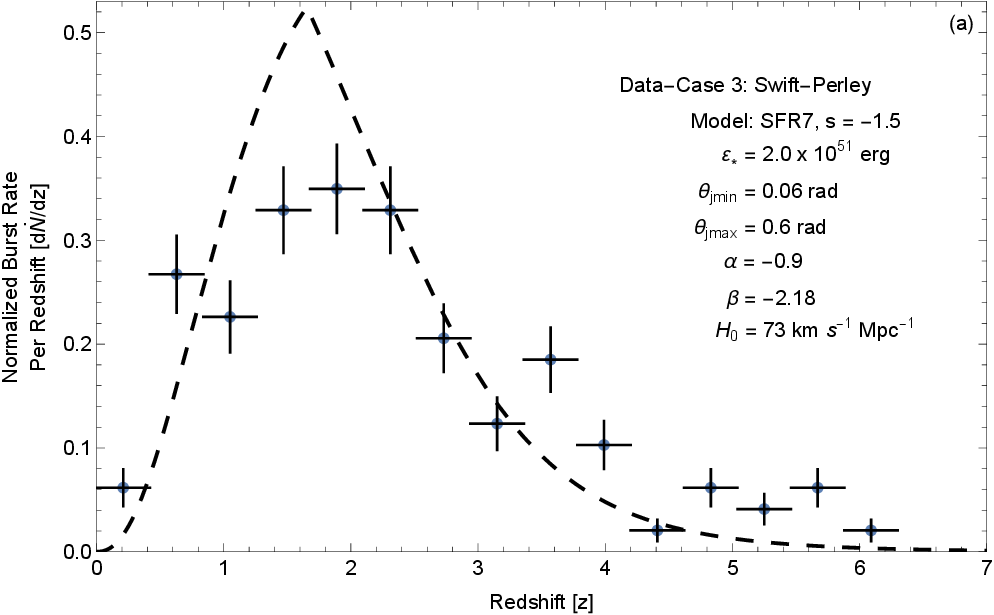}{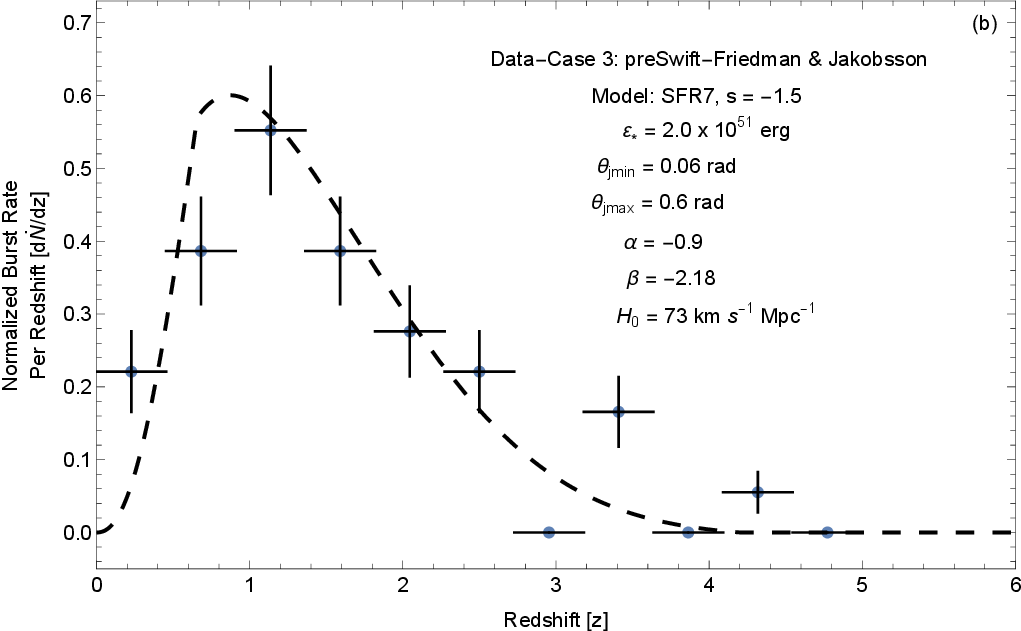}
\caption{\footnotesize Panels ({\it a}) and ({\it b}) show the normalized burst rate per redshift distribution for the~\Swift~and pre\Swift~samples using Case-3 parameter values, analogous to Figures 11({\it a}) and 11({\it b}) but for the SFR7 model.}
\label{fig18} 
\finfig
\begfig[t] \hskip-0.25in \epsscale{1.1} \plottwo{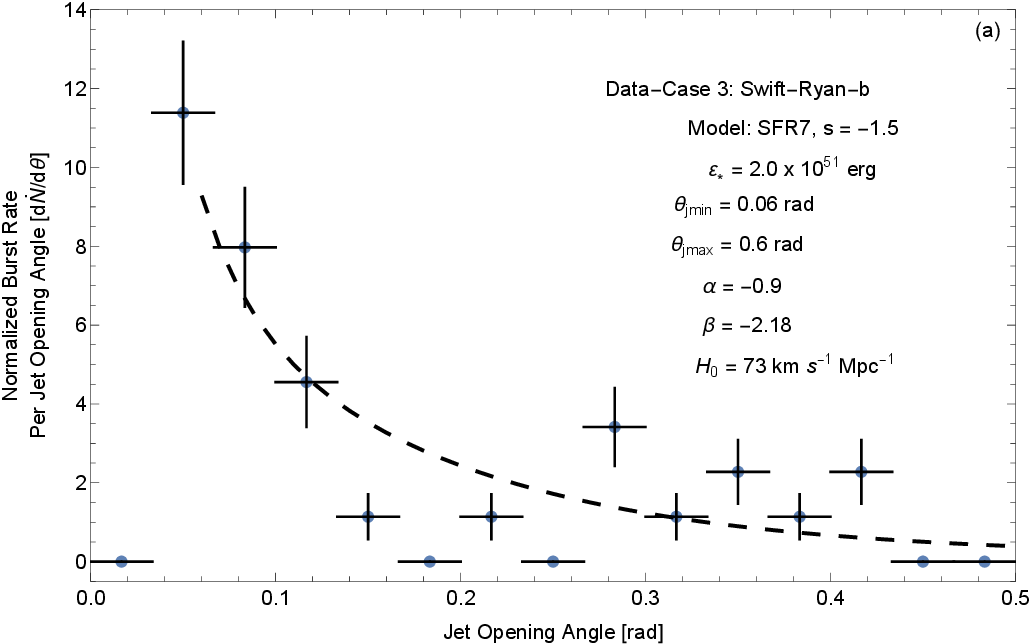}{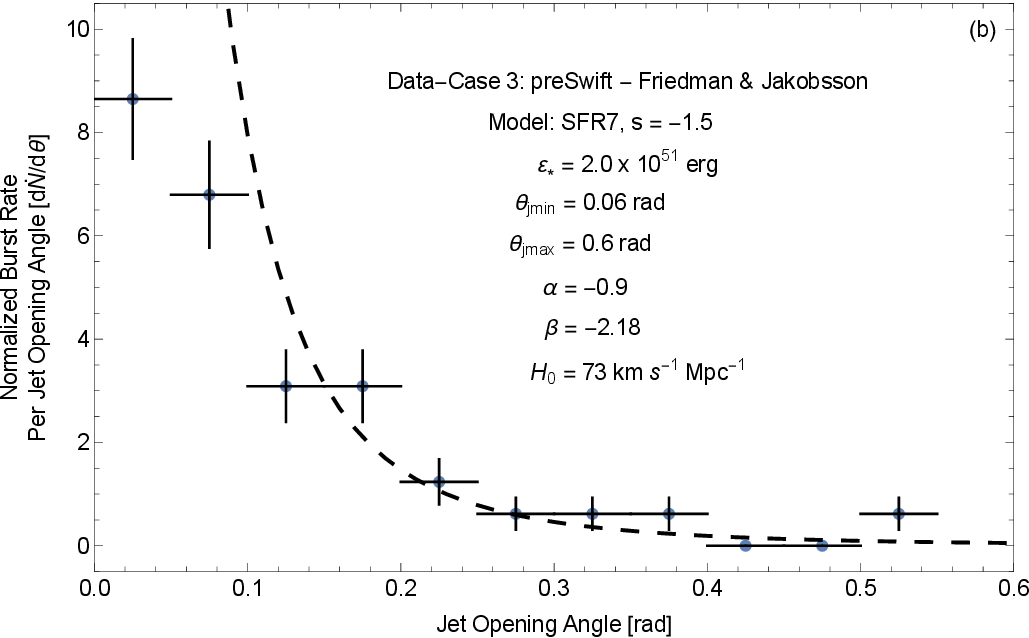}
\caption{\footnotesize Panels ({\it a}) and ({\it b}) show the normalized burst rate per jet opening angle distribution for the~\Swift~and pre\Swift~samples using Case-3 parameter values, analogous to Figures 12({\it a}) and 12({\it b}) but for the SFR7 model.}
\label{fig19} 
\finfig
\begfig[t] \hskip-0.25in \epsscale{1.1} \plottwo{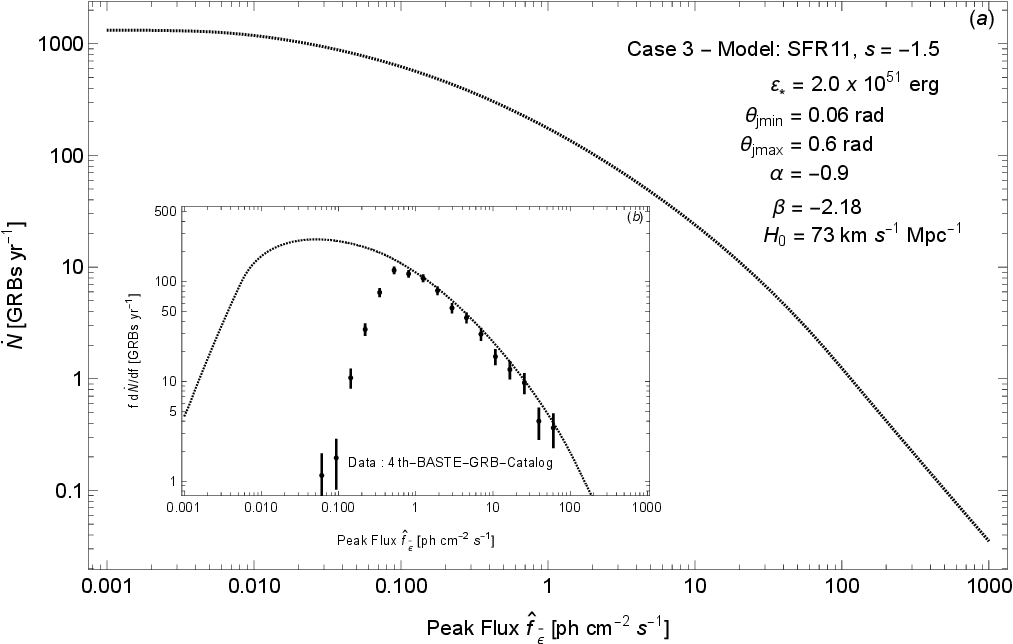}{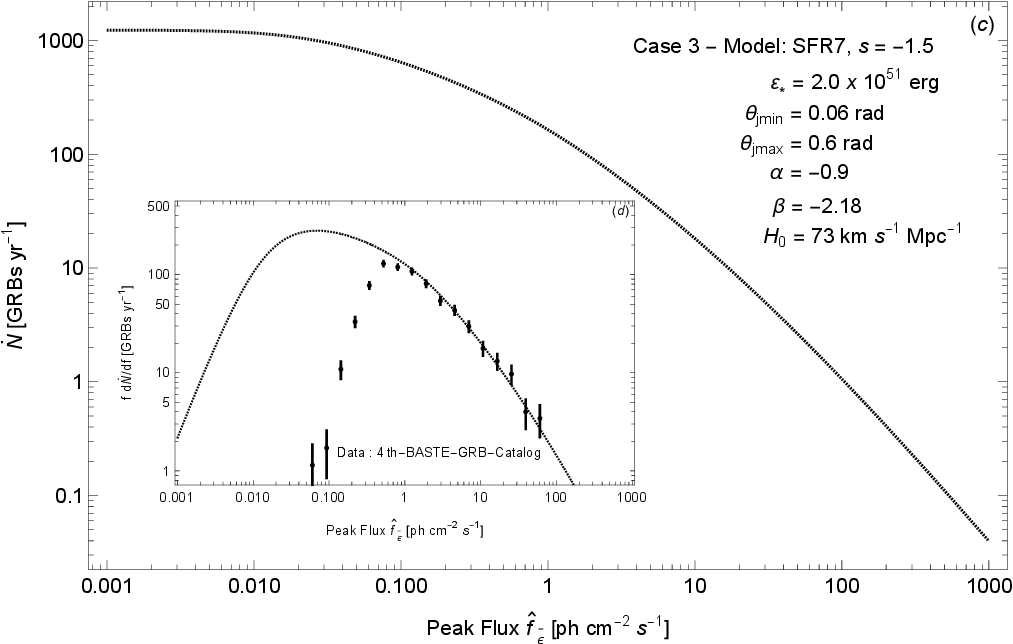}
\caption{\footnotesize Panels ({\it a}) and ({\it c}) are the integral size distributions and the insets are the differential size distributions, similar to Figure-13, but for SFR11 and SFR7, respectively. }
\label{fig20} 
\finfig
%

%
\begfig[t] \hskip-0.25in \epsscale{1.1} \plottwo{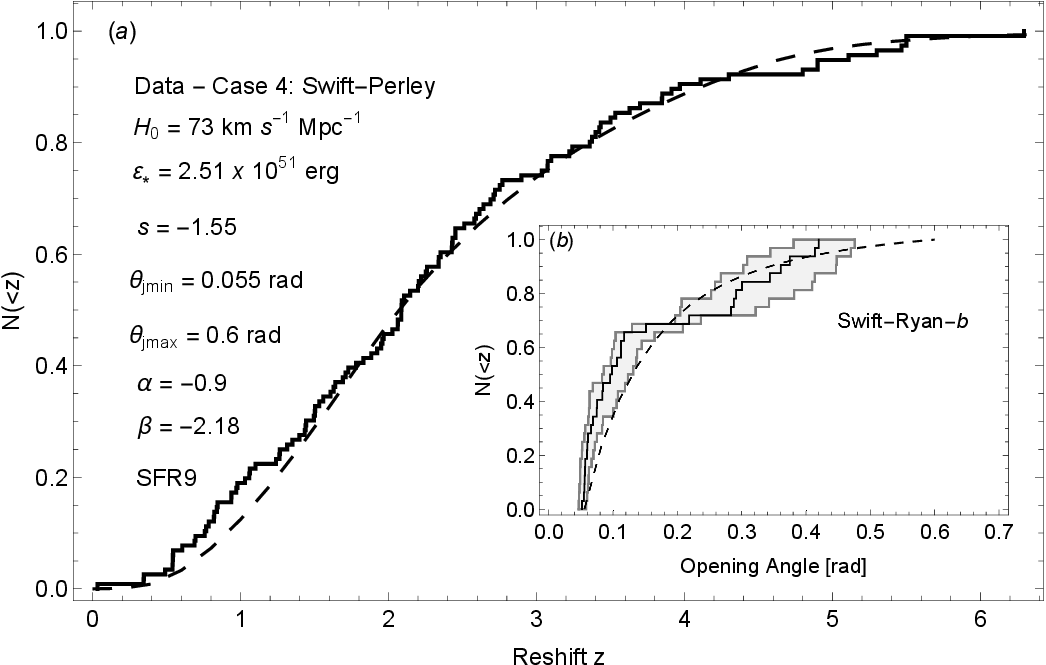}{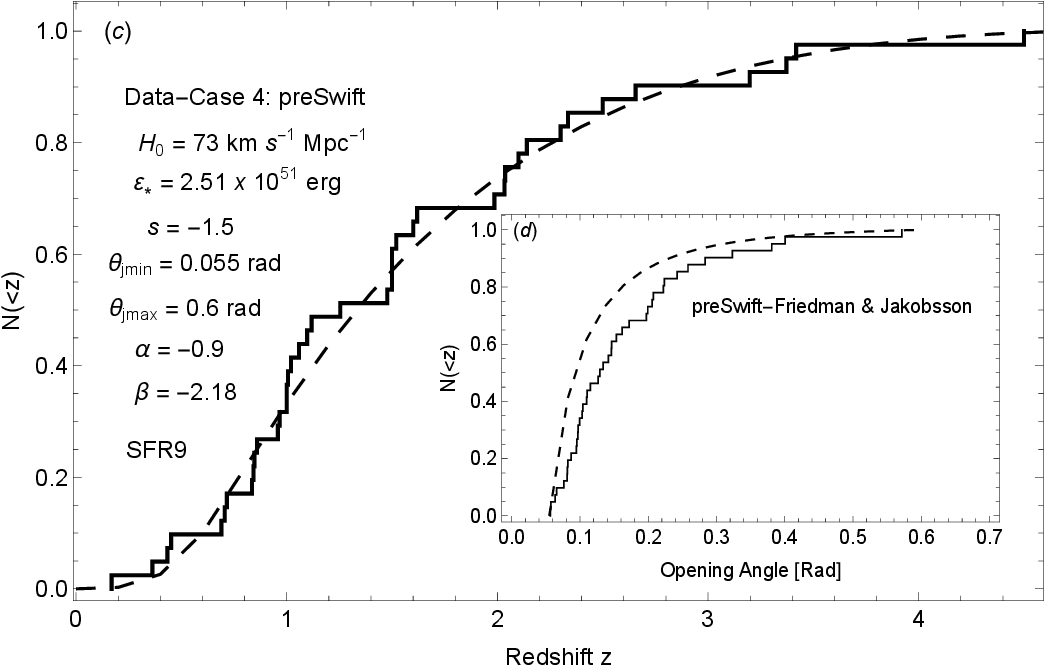}
\caption{\footnotesize Panel ({\it a}) and inset ({\it b}) show the~\Swift~redshift and jet-opening-angle cumulative distributions, while panel ({\it c}) and inset ({\it d}) show the corresponding pre-\Swift~distributions for the Case-4 model. The solid curves represent the observed~\Swift~(Perley; Ryan) and pre\Swift~(Friedman; Jakobsson) samples, with the shaded region in inset ({\it b}) indicating the~\Swift-Ryan uncertainties. The long-dashed curves in all panels show the model predictions assuming SFR9 model.}
\label{fig21} 
\finfig
\begfig[t] \hskip-0.25in \epsscale{1.1} \plottwo{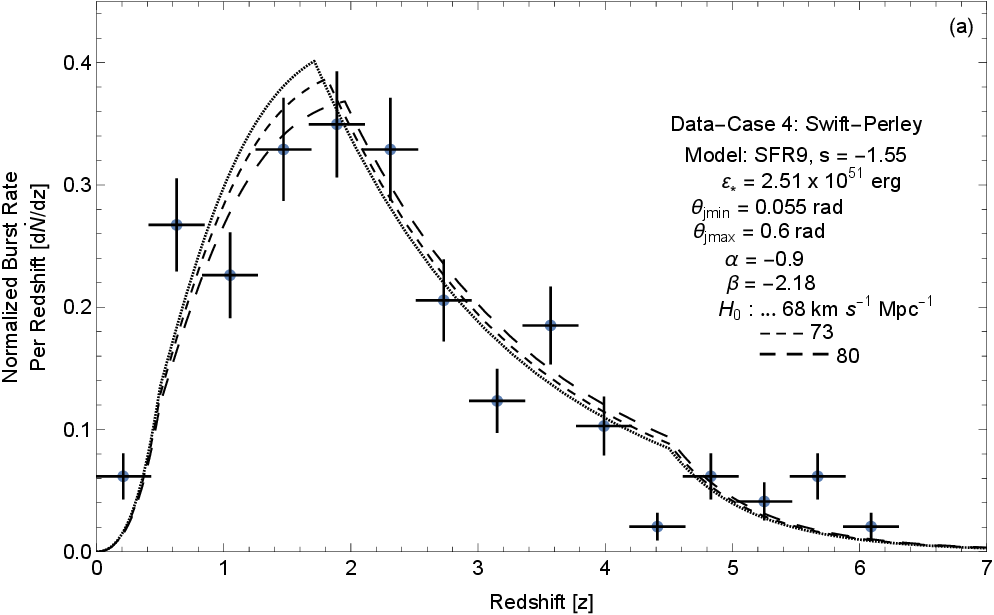}{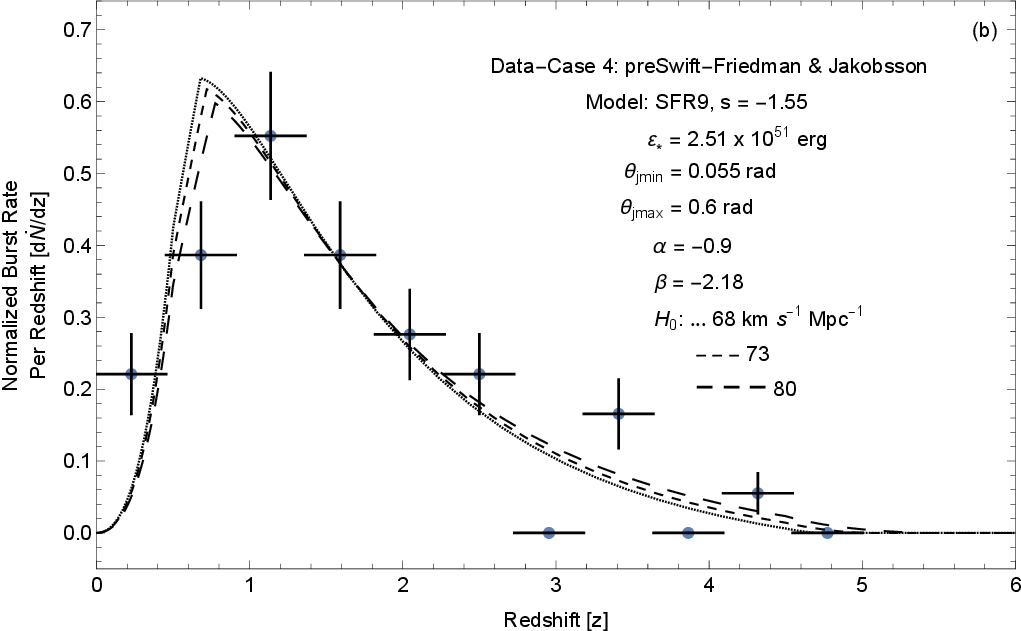}
\caption{\footnotesize ({\it a}) Normalized burst-rate per redshift distribution for the~\Swift-Perley sample and ({\it b}) the pre\Swift-Friedman \& Jakobsson sample. The long-dashed curves in both panels show the model predictions using Case-4 parameter values assuming SFR9 model.}
\label{fig22} 
\finfig
\begfig[t] \hskip-0.25in \epsscale{1.1} \plottwo{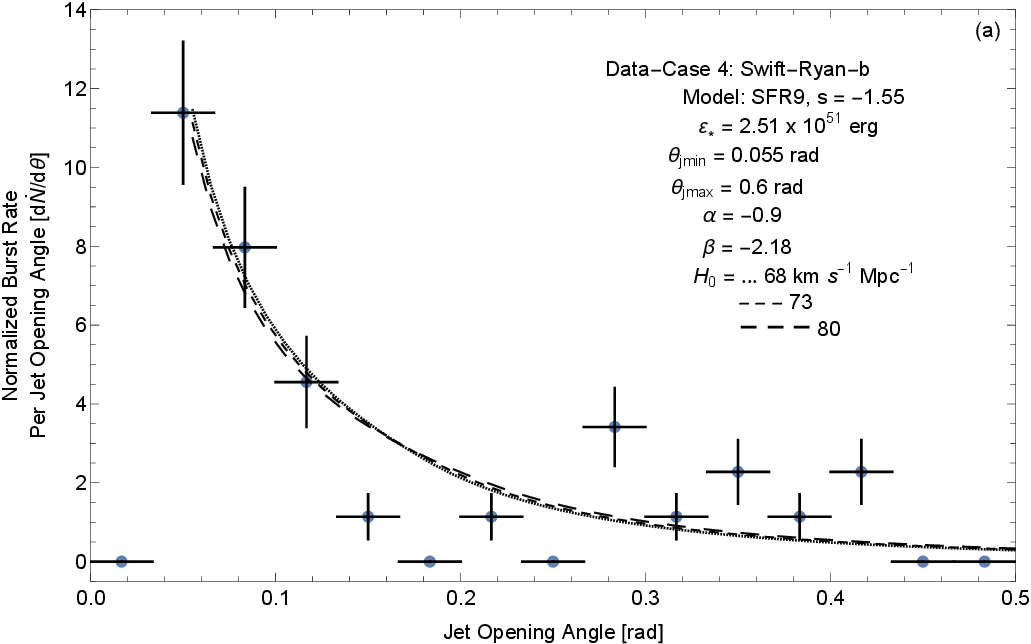}{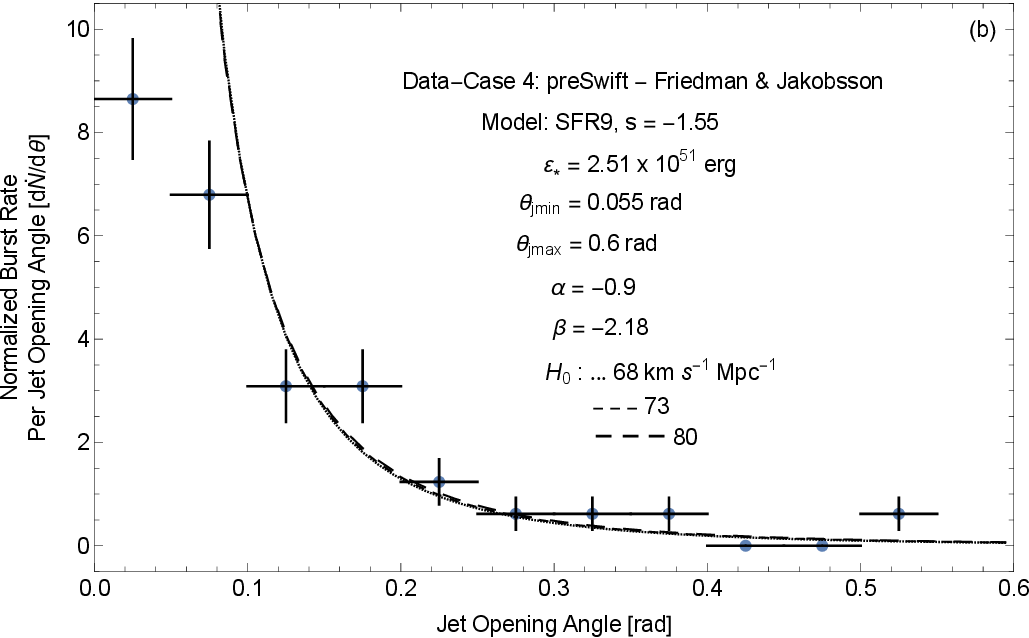}
\caption{\footnotesize ({\it a}) Normalized burst-rate per jet opening angle distribution for the~\Swift-Ryan-b sample and ({\it b}) the pre\Swift-Friedman \& Jakobsson sample. The long-dashed curves in both panels show the model predictions using Case-4 parameter values assuming SFR9 model.}
\label{fig23} 
\finfig
%
\begfig[t] \hskip-0.25in \epsscale{0.8} \plotone{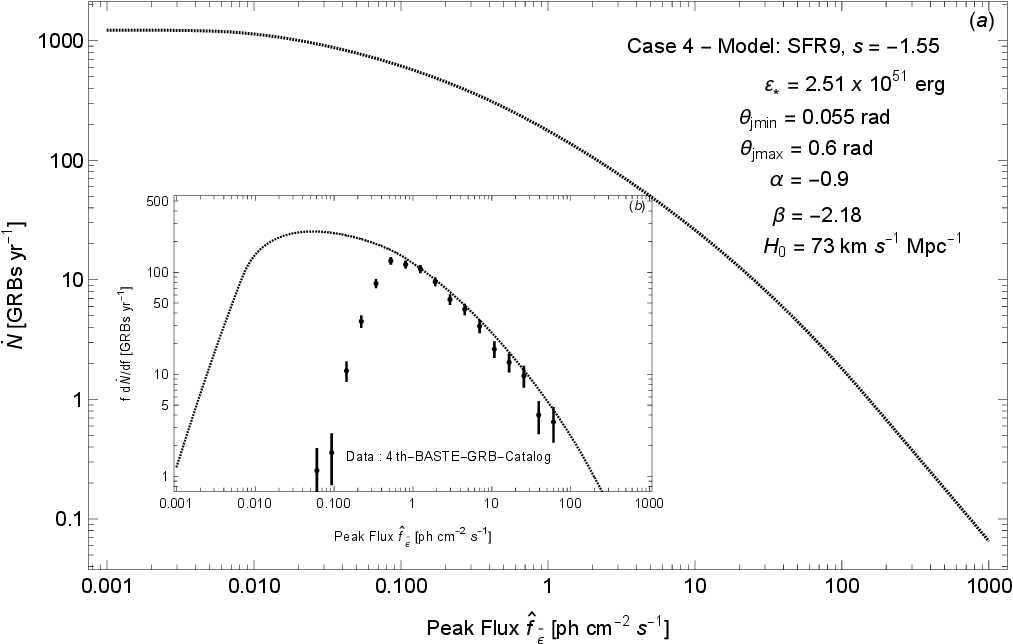}
\caption{\footnotesize ({\it a}) Integral size distributions and ({\it b}) the differential size distributions assuming SFR9 Case-4 model. The filled circles curve represents the 1024 ms trigger-timescale data from the 4B- Catalog, the same as in Figures~\ref{fig9}a and~\ref{fig9}b.}
\label{fig24} 
\finfig

\section{CONCLUSIONS}
\citet{ld07} developed a physical model to understand the differences between the redshift and jet opening angle distributions of \Swift~and pre\Swift~GRBs, taking into account the different detector triggering properties. Their GRB model is based on the uniform jet model and a flat GRB spectrum with an assumption that the LGRB density rate is proportional to the measured SFR. They showed that a consistent agreement was only possible by providing a positive evolution of the SFR history of GRBs to high redshifts. \citet{lm17} revisited the work done by LD07 and performed a timely study of the rate density of GRBs with an assumed broken power-law GRB spectrum. Utilizing more than 100 LGRBs in the \Swift~sample that includes both the observed estimated redshifts and jet opening angles, they obtained a GRB burst rate functional form that provides good agreement to the pre\Swift~and \Swift~redshift and jet opening angle distributions with a GRB density rate, SFR9, that is similar to the \citet{hb06} observed star formation history (SFR7) and as extended by \citet{li08}. However, their worked of redshift model distribution indicates an excess of LGRBs at low redshift below $z \sim 2$ in the \Swift~sample, consistent with \citet{pkk15}, \citet{yu15}, and \citet{laj19}. Using the \Swift-Perley LGRB sample (Swift Gamma-Ray Burst Host Galaxy Legacy Survey) and applying the same physical parameter values adopted for the pre\Swift~and \Swift-Ryan-2012 samples, \citet[][]{lrm20} obtained consistent results with this data set and found no excess of LGRBs at any redshift. Moreover, \citet{lrm20} noticed that the cumulative redshift distribution is only weakly sensitive to variation in the Hubble constant, $\H0$, whereas the differential event-rate-per-redshift distribution shows greater sensitivity. Motivated by this result, in the present work we revisit the analysis of \citet{lrm20} to examine how variations in $\H0$ affect the predicted differential distributions.

Generally, our results show that variations in $\H0$ have only a weak effect on the predicted cumulative and differential redshift and jet opening-angle distributions. For the Case-1 through Case-3 parameter sets, the differences among the $H_0$ models are most apparent near $z \sim 1-2$ in the~\Swift-Perley differential redshift distribution. However, this behavior cannot be attributed uniquely to $\H0$, because the predicted distribution in this redshift range also depends on the adopted GRB formation-rate model and physical parameters, particularly the jet opening-angle distribution index $s$ and the mean absolute emitted gamma-ray energy $\Estarg$. With the Case-4 parameter set and the GRB rate evolution SFR9, the discrepancy between the predicted and observed distributions near $z \sim 1-2$ is no longer apparent, and $H_0 = 68, 73, 80$ km s$^{-1}$ Mpc$^{-1}$ all produce predicted distributions consistent with the observational comparisons considered in this work.

Using the Case-3 and Case-4 parameters, we obtain the mean beaming factors of $< f_b >^{-1} \approx 60$ and $\sim 67$, respectively. These values are consistent with those obtained by \citet{bfk03}, \citet{ggl04}, \citet{gpw05}, and \citet{ggs12}, indicating that the jet geometry adopted in our model is consistent with independent constraints from GRB population studies. Following previous studies of jet-induced luminosity by Le \& Dermer, we infer the apparent isotropic luminosity is $\propto L^{-0.5, -0.45}_*$ within the uniform jet model. Our uniform-jet model induces a much shallower isotropic-equivalent luminosity function than the empirical broken power-law, indicating that the luminosity diversity in our model is primarily geometric rather than intrinsic. Moreover, \citet{pgs15} demonstrated that the apparent isotropic-equivalent luminosity function of long GRBs is strongly shaped by jet geometry and viewing angle effects. They showed that a narrow intrinsic energy distribution combined with realistic jet structures can reproduce the observed Swift luminosity and redshift distributions without invoking intrinsic luminosity evolution. Our results are consistent with this interpretation and further show that a geometry-induced luminosity function derived from the fitted jet opening-angle distribution provides a unified description of~\Swift~and pre\Swift~observations. 

Using the Case-3 and Case-4 parameters together with the GRB rate evolution, SFR9, our model suggest that LGRBs can be detected by~\Swift~to a maximum redshift $z \approx 11 (14)$, and that about 10\% of Swift LGRBs should occur at $z > 4$, in accord with the data shown in Figure 10(a). Our model also predicts that about 5\% of LGRBs should be detected above $z > 4.5$, while fewer than 15\% of LGRBs should be detected at $z \le 1$. Examination of the~\Swift~GRB samples$^1$ between 2013 and 2015 show that fewer than 5\% of LGRBs per year were detected above $z > 5$, approximately $\sim 10\%$ at $z > 4$, and only about 6\% below $z < 1$. These observed fractions are consistent with our model predictions, suggesting that SFR9 provides a plausible representation of the LGRB density rate. We also compare our model peak-flux distribution with BATSE 4B Catalog size distribution and find that the model predicts a~\Swift~detection rate of $\approx 90$ LGRBs per year, consistent with observed~\Swift~rate.

In conclusion, using a geometry-based uniform jet model with a GRB formation rate tied to the observed cosmic star-formation history, we simultaneously reproduce the overall behavior of the cumulative and differential redshift and jet opening-angle distributions of the~\Swift~and pre\Swift~LGRB samples. Our results show that both the cumulative and differential distributions depend only weakly on variations in the Hubble constant, with their behavior also depending on the adopted GRB formation-rate model and physical parameters. Within the physically and observationally consistent parameter sets considered in this work, Hubble constants in the range $\H0 = 68 - 80$ km s$^{-1}$ Mpc$^{-1}$ produce the cumulative, differential, and peak-flux distributions that provide consistent agreement with the pre\Swift-Friedman, \Swift-Perley redshift and~\Swift-Ryan jet opening angle LGRB samples. Thus, the LGRB distributions considered in this work do not strongly discriminate among the values of the Hubble constant.

\begin{acknowledgments}
We wish to thank the anonymous referee for many useful comments and suggestions that greatly improved the paper. TL also thanks Moayad Smaysem for discussions and comments.
\end{acknowledgments}

\section*{Data Availability}
The data underlying this article will be shared on reasonable request
to the corresponding author (TL).



%
%

%
\begin{deluxetable}{llccccc}
\tabletypesize{\small} %
\tablecaption{{\bf GRB formation rate model parameters}
\label{tbl-1}} %
\tablewidth{-0.pt} %
\tablehead{ %
\colhead{$\rm  $} %
&\colhead{$\rm GRB_{DRM}$}%
&\colhead{$\rm \eta_{_1}$}%
&\colhead{$\rm \eta_{_2}$} %
&\colhead{$\rm \eta_{_3}$} %
&\colhead{$\rm \eta_{_4}$} %
&\colhead{$\rm Ref.$} 
} \startdata
Flat $\nu F_{\nu}$ 
& SFR5
& 0.015
& 0.12
& 3.0
& 1.3
& LD07
\\
{ }
&SFR6
& 0.011
& 0.12
& 3.0
& 0.5
& LD07
\\
\hline
Observed star formation rate 
&SFR7$^a$
& 0.0157
& 0.118
& 3.23
& 4.66
& LD07, LM17, LRM20
\\
\hline
{ }
Broken power-law $\nu F_{\nu}$  
&SFR9        
& 4.1
& 0.8        
& -5.1
&  --  
& LM17, LRM20
\\ 
{}
&SFR11
& 5.5
& 0.38
& -4.1
& --
& LRM20
\\
\hline \hline
\enddata
\tablecomments{SFR5, SFR6, SFR9, and SFR11 are the calculated GRB density rate model (DRM). The dash line in the table just means the current model does not have the $\eta_{_4}$ parameter. The last column indicates which paper utilizes the corresponding model. $^a$The parameters values in the SFR7 model are the fitted values to the observed \citet{hb06} SFR history that were discussed in the LD07 paper.}
\end{deluxetable}
\begin{deluxetable}{lcc}
\tablecaption{{\bf Instruments sensitivity summary}
\label{tbl-2}}
\tablehead{\colhead{Instrument} & \colhead{Band (keV)} & \colhead{Sensitivity (erg cm$^{-2}$ s$^{-1}$)}}
\startdata
Swift/BAT & 15--150 & $\sim 1\times10^{-8}$ \\
preSwift/BATSE     & 50--300 & $\sim 1\times10^{-7}$ \\
\enddata
\tablecomments{In the previous work (LD07, LM17, and LRM20) and this present work, we use the above sensitivities for the pre\Swift~and~\Swift~detectors.}
\end{deluxetable}
\begin{deluxetable}{lcccccc}
\tablecaption{{\bf Model Parameters}
\label{tbl-3}}
\tablehead{\colhead{Case} & \colhead{$\Estarg$ (erg)} & \colhead{$\alpha$} & \colhead{$\beta$} & \colhead{$\theta_{\rm j,min} (rad)$} & \colhead{$\theta_{\rm j,max} (rad)$} & \colhead{$s$}}
\startdata
1 & $4.47 \times 10^{51}$ & $-0.9$ & $-2.18$ & $0.04$ & $0.8$ & $-1.55$ \\
2 & $4.47 \times 10^{51}$ & $-0.9$ & $-2.18$ & $0.065$ & $0.8$ & $-1.55$ \\
3 & $2.00 \times 10^{51}$ & $-0.9$ & $-2.18$ & $0.060$ & $0.6$ & $-1.50$ \\
4 & $2.51 \times 10^{51}$ & $-0.9$ & $-2.18$ & $0.055$ & $0.6$ & $-1.55$ \\
\enddata
\tablecomments{Case 1 and Case 2 contain the model parameter values from LRM20. Cases 3 and 4 contain model parameter values that give acceptable fits to both the cumulative and differential distributions in this present work.}
\end{deluxetable}
\begin{deluxetable}{|lcccc||lc|}
\tabletypesize{\small} %
\tablecaption{{\bf The Friedman and the Jakobsson samples}
\label{tbl-4}} %
\tablewidth{-0.pt} %
\tablehead{ %
\colhead{$\rm GRB^a$} %
&\colhead{$\rm z$}%
&\colhead{$\rm \theta_{jet} (deg)$} %
&\colhead{$\rm \alpha$} %
&\colhead{$\rm \beta$ }  %
&\colhead{$\rm GRB^b$} %
&\colhead{$\rm z$}%
} \startdata
970508
& 0.8349
& 21.83
& -1.71				
& -2.20
& 050315
&1.95
\\
970828
& 0.9578
& 7.26
& -0.70				
& -2.07
& 050318
& 1.44
\\
971214        
& 3.418 
& 5.48        
& -0.76				  
& -2.70  
& 050319
& 3.24
\\ 
980326 
& 1
& 6.33
& -1.23				
& -2.48
& 050401
& 2.90
\\
980519
& 2.5
& 3.65
& -1.35				
& -2.30
& 050416A
& 0.65
\\
980613
& 1.0969								
& 12.82
& -1.43				
& -2.70
& 050505
& 4.27
\\
980703 
& 0.9662				
& 11.42
& -1.31				
& -2.40
& 050525
& 0.61
\\
981226
& 1.5				
& 14.80
& -1.25				
& -2.60
& 050603
& 2.82
\\
990123
&1.6004				
& 4.68
& -0.89				
& -2.45
& 050730
& 3.97
\\
990510 
& 1.6187				
& 3.77
& -1.23				
& -2.70
& 050802
& 1.71
\\
990705
& 0.8424				
& 5.5
& -1.05				
& -2.20
& 050814
& 5.3
\\
990712
& 0.4331				
& 11.78
& -1.88				
& -2.48
& 050820A
& 2.61
\\
991208
& 0.7055				
& 8.39
& --
& --
& 050824
& 0.83
\\
991216
& 1.0200				
& 4.66
& -1.23				
& -2.18
& 050904
& 6.29
\\
000131
& 4.5				
& 4.71
& -1.20				
& -2.40
& 050908
& 3.34
\\
000210
& 0.8463				
& 5.43
& --
& --
& 050922B
& 2.20
\\ \cline{6-7}
000301C
& 2.0335				
& 13.88
& --
& --
\\
000418
&1.1182				
&22.95
& --
& --
\\
000630
& 1.5				
& 9.85
& --
& --
\\
000911
& 1.0585				
& 5.58
& -1.11				
& -2.32
\\
000926
& 2.0369				
& 6.28
& --
&--
\\
010222
& 1.4769				
& 3.29
& -1.35				
& -1.64
\\ 
010921
& 0.4509				
& 32.76
& -1.55				
& -2.30
\\
011121
& 0.3620				
& 16.24
& -1.42				
& -2.30
\\
011211
& 2.14				
& 5.98
& -0.84				
& -2.30
\\
020124
& 3.198				
& 11.3
& -0.79				
& -2.30
\\
020405
& 0.6899				
& 7.68
& --
& --
\\
020427
& 2.30				
&18.52
& -1.00				
& -2.10
\\
020813
& 1.254				
& 3.24
& -0.94				
& -1.57
\\
021004
& 2.3351				
& 12.73
& -1.01				
& -2.30
\\
021211
& 1.0060				
& 8.78
& -0.86				
& -2.18
\\
030226 
&1.9860				
& 4.99
& -0.89				
& -2.30 
\\
030323
& 3.3718				
& 5.71
& -1.62				
& -2.30
\\
030328
&1.52				
& 4.37
& -1.14				
& -2.09
\\ 
030329
& 0.1685				
& 6.6
& -1.26				
& -2.28  
\\
030429
& 2.6564				
& 7.41
& -1.12				
& -2.30
\\
030528 
&1				
& 9.27
& -1.33				
& -2.65
\\
030723 
& 2.1				
& 11.89
& -1.00				
& -1.90
\\
040511
& 1.5				
& 5.91
& -0.67				
& -2.30
\\
040924
& 0.8590				
& 8.36
& -1.17				
& -2.30
\\
041006
& 0.7160				
& 8.11
& -1.37				
& -2.30
\\
\enddata
\tablecomments{$^a$These are the pre\Swift~data from the \citet{fb05}.
$^b$These are the \Swift~data from the \citet{jak06}.}
\end{deluxetable}
%

\begin{deluxetable}{|lc|lc|lc|}
\tabletypesize{\small} %
\tablecaption{{\bf  The \Swift-Perley samples}
\label{tbl-5}} %
\tablewidth{-0.pt} %
\tablehead{ %
\colhead{$\rm GRB^{a}$} %
&\colhead{$\rm z$}%
&\colhead{$\rm GRB^{a}$} %
&\colhead{$\rm z$}%
&\colhead{$\rm GRB^{a}$} %
&\colhead{$\rm z$}%
} \startdata
050128	
&5.5	
&070223
&1.6295
&090417B
&0.345
\\
050315
&1.95
&070306
&1.4959
&090418A
&1.608
\\
050318
&1.4436
&070318
&0.840
&090424
&0.544
\\
050319
&3.2425
&070328
&2.0627
&090516A
&4.109
\\
050401
&2.8983
&070419B
&1.9588
&090519
&3.85
\\
050525A
&0.606
&070508
&0.82
&090530
&1.266
\\
050726
&3.5
&070521
&2.0865
&090618
&0.54
\\
050730
&3.9693
&070621
&5.5
&090709A
&1.8
\\
050802
&1.7102
&070721B
&3.6298
&090715B
&3.00
\\
050803
&4.3
&070808
&1.35
&090812
&2.452
\\
050814
&5.3
&071020
&2.1462
&090814A
&0.696
\\
050820A
&2.6147
&071021
&2.4520
&090926B
&1.24
\\
050822
&1.434
&071025
&4.8
&091018
&0.971
\\
050904
&6.295
&071112C
&0.8227
&091029
&2.752
\\
050922B
&4.9
&080205
&2.72
&091109A
&3.076
\\
050922C
&2.1995
&080207
&2.0858
&091127
&0.490
\\
051001
&2.4296
&080210
&2.6419
&091208B
&1.0633
\\
051006
&1.059
&080310
&2.4274
&100305A
&--
\\
060115
&3.5328
&080319A
&2.0265
&100615A
&1.398
\\ 
060202
&0.785
&080319B
&0.9382
&100621A
&0.542
\\ 
060204B
&2.3393
&080325
&1.78
&100728B
&2.106
\\
060210
&3.9122
&080411
&1.0301
&100802A
&3.1
\\
060218
&0.0331
&080413A
&2.4330
&100814A
&1.44
\\
060306
&1.559
&080413B
&1.1014
&110205A
&2.22
\\
060502A
&1.5026
&080430
&0.767
&110709B
&2.09
\\
060510B
&4.9
&080603B
&2.6892
&120119A
&1.728
\\
060522
&5.11
&080605
&1.6403
&120308A
&3.7
\\ \cline{5-6}
060526
&3.2213
&080607
&3.0368
\\
060607A
&3.0749
&080710
&0.8454
\\
060707
&3.4240
&080721
&2.5914
\\
060714
&2.7108
&080804
&2.2045
\\
060719
&1.5320
&080805
&1.5042
\\
060729
&0.5428
&080810
&3.3604
\\
060814
&1.9229
&080916A
&0.6887
\\
060908
&1.8836
&080928
&1.6919
\\
060912A
&0.937
&081008
&1.967
\\
060927
&5.467
&081029
&3.8479
\\
061007
&1.2622
&081109A
&0.9787
\\
061021
&0.3463
&081118
&2.58
\\
061110A
&0.7578
&081121
&2.512
\\
061110B
&3.4344
&081128
&3.4
\\
061121
&1.3145
&081210
&2.0631
\\
061202
&2.253
&081221
&2.26
\\
061222A
&2.088
&081222
&2.77
\\
070110
&2.3521
&090313
&3.375
\\
070129
&2.3384
&090404
&3.0
\\
\enddata

\tablecomments{$^a$These are samples from the \citet{per16}.}
\end{deluxetable}
%

\begin{deluxetable}{|lcc|lcc|lcc|}
\tabletypesize{\small} %
\tablecaption{{\bf \Swift-Ryan sample}
\label{tbl-6}} %
\tablewidth{-0.pt} %
\tablehead{ %
\colhead{$\rm GRB$} %
&\colhead{$\rm z$}%
&\colhead{$\rm \theta_{jet} (rad)$} %
&\colhead{$\rm GRB$} %
&\colhead{$\rm z$}%
&\colhead{$\rm \theta_{jet} (rad)$} %
&\colhead{$\rm GRB$} %
&\colhead{$\rm z$}%
&\colhead{$\rm \theta_{jet} (rad)$} %
} \startdata
121024A$^a$	
& 2.298	
& 0.0565$^{+0.0179}_{-0.008}$
&090328A	$^b$
&0.736	
&0.32$^{+0.13}_{-0.17}$	
&061007$^b$	
&1.2622	
&0.31$^{+0.13}_{-0.17}$	
\\
120922A$^b$	
& 3.1	
&0.3$^{+0.14}_{-0.13}$	
&090313$^b$	
&3.375	
&0.127$^{+0.143}_{-0.04}$	
&061006$^b$	
&0.4377	
&0.407$^{+0.068}_{-0.173}$	
\\
120909A$^b$	
&3.93	
&0.22$^{+0.18}_{-0.13}$	
&090205$^a$	
&4.6497	
&0.0513$^{+0.0081}_{-0.0046}$	
&060926$^b$	
&3.2086	
&0.29$^{+0.15}_{-0.16}$	
\\
120802A$^b$	
&3.796	
&0.21$^{+0.18}_{-0.15}$	
&090113$^b$	
&1.7493	
&0.3$^{+0.13}_{-0.14}$	
&060912A$^b$	
&0.937	
&0.227$^{+0.152}_{-0.07}$	
\\
120712A$^b$	
&4.1745	
&0.301$^{+0.125}_{-0.099}$	
&081222$^b$	
&2.77	
&0.0844$^{+0.0081}_{-0.0120}$	
&060904B$^{a,b}$	
&0.7029	
&0.083$^{+0.048}_{-0.015}$	
\\
120404A$^b$	
&2.876	
&0.27$^{+0.15}_{-0.15}$	
&081221$^b$	
&2.26	
&0.34$^{+0.11}_{-0.093}$	
&060714$^a$	
&2.7108	
&0.0557$^{+0.0104}_{-0.0072}$	
\\
120119A$^b$	
&1.728	
&0.0510$^{+0.0481}_{-0.0046}$	
&081203A$^b$	
&2.1	
&0.121$^{+0.058}_{-0.063}$	
&060708$^a$	
&1.92	
&0.0597$^{+0.0115}_{-0.008}$	
\\
120118B$^b$	
&2.943	
&0.29$^{+0.14}_{-0.15}$	
&081109$^b$	
&0.9787	
&0.26$^{+0.16}_{-0.14}$	
&060707$^b$	
&3.424	
&0.22$^{+0.19}_{-0.16}$	
\\
111211A$^b$ 
&0.478	
&0.3$^{+0.14}_{-0.15}$	
&081029$^b$	
&3.8479	
&0.1615$^{+0.0055}_{-0.0061}$	
&060614$^b$	
&0.1257	
&0.293$^{+0.122}_{-0.085}$	
\\
111209A$^b$	
&0.677	
&0.34$^{+0.11}_{-0.13}$	
&081028$^b$	
&3.038	
&0.3$^{+0.13}_{-0.15}$	
&060607A$^{a,b}$	
&3.0749	
&0.374$^{+0.095}_{-0.072}$	
\\
111107A$^b$	
&2.893	
&0.28$^{+0.15}_{-0.15}$	
&081007$^b$	
&0.5295	
&0.159$^{+0.183}_{-0.066}$	
&060605$^b$	
&3.773	
&0.04614$^{+0.18939}_{-0.00096}$	
\\
110918A$^b$	
&0.982	
&0.35$^{+0.11}_{-0.17}$	
&080928$^b$	
&1.6919	
&0.25$^{+0.16}_{-0.17}$	
&060522$^b$	
&5.11	
&0.3$^{+0.13}_{-0.14}$	
\\
110818A$^b$	
&3.36	
&0.24$^{+0.18}_{-0.16}$	
&080913$^b$	
&6.7	
&0.359$^{+0.099}_{-0.125}$	
&060319$^b$	
&1.172	
&0.32$^{+0.12}_{-0.14}$	
\\
110808A$^b$	
&1.348	
&0.19$^{+0.19}_{-0.13}$	
&080810$^b$	
&3.3604	
&0.34$^{+0.11}_{-0.27}$	
&060218$^b$	
&0.0331	
&0.33$^{+0.12}_{-0.16}$	
\\
110801A$^{a,b}$	
&1.858	
&0.419$^{+0.056}_{-0.075}$	
&080721$^{a,b}$	
&2.5914	
&0.1117$^{+0.0109}_{-0.0083}$	
&060210$^a$	
&3.9122	
&0.0604$^{+0.0236}_{-0.0093}$	
\\
110715A$^b$	
&0.82	
&0.3$^{+0.14}_{-0.17}$	
&080607$^b$	
&3.0368	
&0.26$^{+0.16}_{-0.14}$	
&060206$^b$	
&4.059	
&0.377$^{+0.084}_{-0.111}$	
\\
110503A$^{a,b}$	
&1.613	
&0.1057$^{+0.0302}_{-0.0079}$	
&080605$^{a,b}$	
&1.6403	
&0.36$^{+0.087}_{-0.108}$	
&060202$^b$	
&0.785	
&0.28$^{+0.15}_{-0.15}$	
\\
110422$^{a,b}$	
&1.77	
&0.075$^{+0.025}_{-0.012}$	
&080603A$^b$	
&1.6880	
&0.31$^{+0.13}_{-0.16}$	
&060123$^b$	
&0.56	
&0.32$^{+0.13}_{-0.15}$	
\\
110213A$^b$	
&1.46	
&0.29$^{+0.14}_{-0.15}$	
&080430$^b$	
&0.767	
&0.0553$^{+0.0052}_{-0.0079}$	
&051111$^b$	
&1.55	
&0.28$^{+0.15}_{-0.15}$	
\\
110205A$^b$	
&2.22	
&0.388$^{+0.077}_{-0.145}$	
&080413B$^a$	
&1.1014	
&0.117$^{+0.02}_{-0.021}$	
&051109B$^b$	
&0.08	
&0.06$^{+0.162}_{-0.012}$	
\\
110128A$^b$	
&2.339	
&0.469$^{+0.023}_{-0.047}$	
&080413A$^b$	
&2.433	
&0.26$^{+0.16}_{-0.15}$	
&051016B$^b$	
&0.9364	
&0.35$^{+0.11}_{-0.24}$	
\\
100906A$^a$	
&1.727	
&0.054$^{+0.0038}_{-0.0079}$	
&080411$^b$	
&1.0301	
&0.097$^{+0.362}_{-0.017}$	
&051006$^b$	
&1.059	
&0.29$^{+0.15}_{-0.15}$	
\\
100901A$^a$	
&1.408	
&0.413$^{+0.033}_{-0.033}$	
&080319B$^{a,b}$	
&0.9382	
&0.098$^{+0.046}_{-0.013}$	
&050922C$^{a,b}$	
&2.1995	
&0.074$^{+0.033}_{-0.011}$	
\\
100728B$^b$	
&2.106	
&0.26$^{+0.15}_{-0.17}$	
&080310$^b$	
&2.4274	
&0.097$^{+0.263}_{-0.052}$	
&050915A	$^b$
&2.5273	
&0.24$^{+0.18}_{-0.14}$	
\\
100728A$^{a,b}$	
&1.567	
&0.112$^{+0.052}_{-0.011}$	
&080210$^b$	
&2.6419	
&0.164$^{+0.147}_{-0.098}$	
&050908$^b$	
&3.3467	
&0.26$^{+0.14}_{-0.13}$	
\\
100621A$^b$	
&0.542	
&0.0477$^{+0.0047}_{-0.002}$	
&080207$^b$	
&2.0858	
&0.31$^{+0.13}_{-0.14}$	
&050824$^b$	
&0.8278	
&0.140$^{+0.229}_{-0.088}$	
\\
100513A$^b$	
&4.8	
&0.28$^{+0.15}_{-0.15}$	
&080129$^b$	
&4.349	
&0.19$^{+0.2}_{-0.12}$	
&050820A$^{a,b}$	
&2.6147	
&0.151$^{+0.057}_{-0.022}$	
\\
100425A$^b$	
&1.755	
&0.16$^{+0.22}_{-0.10}$	
&071122$^b$	
&1.14	
&0.25$^{+0.16}_{-0.16}$	
&050801$^b$	
&1.38	
&0.29$^{+0.14}_{-0.12}$	
\\
100424A$^b$	
&2.465	
&0.17$^{+0.22}_{-0.11}$	
&071112C$^b$	
&0.8227	
&0.398$^{+0.071}_{-0.117}$	
&050730$^b$	
&3.9693	
&0.108$^{+0.301}_{-0.058}$	
\\
100418A$^b$	
&0.6235	
&0.23$^{+0.16}_{-0.12}$	
&071031$^b$	
&2.6918	
&0.3$^{+0.13}_{-0.15}$	
&050525A$^{a,b}$	
&0.606	
&0.0551$^{+0.0069}_{-0.0062}$	
\\
100302A$^b$	
&4.813	
&0.19$^{+0.20}_{-0.13}$	
&071025$^b$	
&5.2	
&0.3$^{+0.14}_{-0.13}$	
&050505$^b$	
&4.27	
&0.216$^{+0.09}_{-0.156}$	
\\
091208B$^a$	
&1.0633	
&0.0952$^{+0.0098}_{-0.0127}$	
&071010B$^b$	
&0.947	
&0.34$^{+0.12}_{-0.22}$	
&050408$^b$	
&1.2356	
&0.29$^{+0.14}_{-0.16}$	
\\
091127$^b$	
&0.49	
&0.151$^{+0.193}_{-0.071}$	
&071003$^b$	
&1.6044	
&0.3$^{+0.14}_{-0.18}$	
&050315$^{a,b}$	
&1.95	
&0.343$^{+0.038}_{-0.035}$	
\\
091029$^b$	
&2.752	
&0.21$^{+0.19}_{-0.13}$	
&070810A$^b$	
&2.17	
&0.104$^{+0.196}_{-0.049}$	
&050219A$^b$	
&0.2115	
&0.28$^{+0.15}_{-0.16}$	
\\ \cline{7-9}
091024$^b$	
&1.092	
&0.31$^{+0.13}_{-0.15}$	
&070721B$^a$	
&3.6298	
&0.084$^{+0.034}_{-0.029}$	
\\ 
091020$^b$	
&1.71	
&0.28$^{+0.15}_{-0.13}$	
&070714B$^b$	
&0.923	
&0.33$^{+0.11}_{-0.11}$	
\\
091018$^b$	
&0.971	
&0.30$^{+0.14}_{-0.16}$	
&070611$^b$	
&2.0394	
&0.27$^{+0.15}_{-0.11}$	
\\
091003$^b$	
&0.8969	
&0.31$^{+0.13}_{-0.17}$	
&070529$^b$	
&2.4996	
&0.25$^{+0.16}_{-0.12}$	
\\
090902B$^b$ 
&1.822	
&0.31$^{+0.13}_{-0.16}$	
&070521$^b$	
&1.7	
&0.154$^{+0.161}_{-0.079}$	
\\
090812$^b$	
&2.452	
&0.29$^{+0.14}_{-0.15}$	
&070506$^b$	
&2.309	
&0.25$^{+0.17}_{-0.16}$	
\\
090809$^b$	
&2.737	
&0.29$^{+0.13}_{-0.12}$	
&070419A$^b$	
&0.97	
&0.18$^{+0.19}_{-0.11}$	
\\
090726$^b$	
&2.71	
&0.183$^{+0.195}_{-0.094}$	
&070318$^b$	
&0.84	
&0.3$^{+0.11}_{-0.12}$	
\\
090709A$^a$	
&1.8	
&0.287$^{+0.126}_{-0.091}$	
&070306$^a$	
&1.49594	
&0.291$^{+0.03}_{-0.033}$	
\\
090618$^{a,b}$	
&0.54	
&0.059$^{+0.0025}_{-0.0026}$	
&070110$^b$	
&2.3521	
&0.33$^{+0.12}_{-0.15}$	
\\
090519$^b$	
&3.85	
&0.27$^{+0.15}_{-0.16}$	
&070103$^b$	
&2.6208	
&0.29$^{+0.15}_{-0.15}$	
\\
090516A$^a$	
&4.109	
&0.0656$^{+0.0035}_{-0.0045}$	
&061222A$^a$	
&2.088	
&0.0684$^{+0.0142}_{-0.0064}$	
\\
090424$^a$	
&0.544	
&0.218$^{+0.018}_{-0.014}$	
&061126$^{a,b}$	
&1.159	
&0.283$^{+0.126}_{-0.077}$	
\\
090417B$^b$	
&0.345	
&0.183$^{+0.232}_{-0.074}$		
&061121$^b$	
&1.3145	
&0.094$^{+0.118}_{-0.03}$	
\\
090407$^a$	
&1.4485	
&0.3$^{+0.042}_{-0.033}$		
&061021$^b$	
&0.3463	
&0.133$^{+0.06}_{-0.021}$	
\\
\enddata

\tablecomments{$^a$ These are a subsample between 2005 and 2012 (\Swift-Ryan-b) from \citet{rya15}. \\
$^b$These are samples between 2005 and 2012 (\Swift-Ryan 2012) from \citet{rya15}.}
\end{deluxetable}
%





\end{document}